\documentclass[twocolumn,prl,showpacs,superscriptaddress,amsmath,amssymbl]{revtex4-1}

\usepackage{amssymb}
\usepackage{graphicx}
\usepackage{dcolumn}
\usepackage{bm}
\usepackage{color}

\hfuzz=\maxdimen
\begin{document}

\title{RbEu(Fe$_{1-x}$Ni$_x$)$_4$As$_4$: From a ferromagnetic superconductor to a superconducting ferromagnet}

\author{Yi Liu}
\thanks{These authors contribute equally to this work.}
\affiliation{Department of Physics, Zhejiang University, Hangzhou 310027, China}

\author{Ya-Bin Liu}
\thanks{These authors contribute equally to this work.}
\affiliation{Department of Physics, Zhejiang University, Hangzhou 310027, China}

\author{Ya-Long Yu} \affiliation{Department of Physics, Zhejiang University, Hangzhou 310027, China}


\author{Qian Tao} \affiliation{Department of Physics, Zhejiang University, Hangzhou 310027, China}

\author{Chun-Mu Feng} \affiliation{Department of Physics, Zhejiang University, Hangzhou 310027, China}

\author{Guang-Han Cao} \email[]{ghcao@zju.edu.cn}
\affiliation{Department of Physics, Zhejiang University, Hangzhou
310027, China} \affiliation{State Key Lab of Silicon Materials,
Zhejiang University, Hangzhou 310027, China} \affiliation{Collaborative Innovation Centre of Advanced Microstructures, Nanjing 210093, China}

\date{\today}



\begin{abstract}
The intrinsically hole-doped RbEuFe$_4$As$_4$ exhibits bulk superconductivity at $T_{\mathrm{sc}}=36.5$ K and ferromagnetic ordering in the Eu sublattice at $T_\mathrm{m}=15$ K. Here we present a hole-compensation study by introducing extra itinerant electrons via a Ni substitution in the ferromagnetic superconductor RbEuFe$_4$As$_4$ with $T_{\mathrm{sc}}>T_{\mathrm{m}}$. With the Ni doping, $T_{\mathrm{sc}}$ decreases rapidly, and the Eu-spin ferromagnetism and its $T_{\mathrm{m}}$ remain unchanged. Consequently, the system RbEu(Fe$_{1-x}$Ni$_x$)$_4$As$_4$ transforms into a superconducting ferromagnet with $T_{\mathrm{m}}>T_{\mathrm{sc}}$ for $0.07\leq x\leq0.08$. The occurrence of superconducting ferromagnets is attributed to the decoupling between Eu$^{2+}$ spins and superconducting Cooper pairs. The superconducting and magnetic phase diagram is established, which additionally includes a recovered yet suppressed spin-density-wave state.

\end{abstract}

\maketitle

\section{\label{sec:level1}Introduction}

Superconductivity (SC) and ferromagnetism (FM) are basically antagonistic and incompatible~\cite{buzdin1985,buzdin2005}, and only in a very few cases can they coexist simultaneously in a single material~\cite{buzdin1985,cgh2012}. The relative robustness of SC and FM can be reflected by the superconducting critical temperature $T_\mathrm{sc}$ and the (ferro)magnetic transition temperature $T_\mathrm{m}$. Materials with $T_\mathrm{sc} > T_\mathrm{m}$ were earlier called ``ferromagnetic superconductors" (FSCs)~\cite{buzdin1985}, and those with $T_\mathrm{m} > T_\mathrm{sc}$ were then termed ``superconducting ferromagnets" (SFMs)~\cite{Felner,chucw,note1}. Generally, a ferromagnetic exchange field prevails over the intrinsic superconducting upper critical field $H_{\mathrm{c2}}^*$ for $T_\mathrm{m} \leq T_\mathrm{sc}$, hence SFMs are particularly rare. So far, examples of SFMs only include U-based germanides with spin-triplet SC~\cite{Mineev} and ruthenocuprates with spin-singlet high-temperature SC~\cite{Bernhard}, the latter of which actually exhibit the coexistence with a canted antiferromagnetism. Note that the classification into FSCs and SFMs is meaningful for studying the way of coexistence of the two antagonistic phenomena~\cite{chucw,SVP.jiao}.

In recent years, Eu-containing 122-type iron arsenides have earned a lot of interest owing to the intriguing interplay between SC and FM~\cite{cgh2012,dressel}. The crystal structure allows the magnetic Eu-atomic planes away from the superconductively active Fe-atom sheets. In non-doped EuFe$_2$As$_2$, the Eu sublattice is of an $A$-type antiferromagnetism below $\sim$19 K while the Fe sublattice exhibits a spin-density wave (SDW) order below $\sim$190 K~\cite{ren2008,js-njp,rxs2009,nd2009}. SC at $T_\mathrm{sc}=$ 20-30 K can be induced by the chemical doping with P~\cite{ren2009}, Co~\cite{jiang2009}, Ru~\cite{jiao2011}, Ir~\cite{jiao2013,hossian2013}, etc. Simultaneously, the Eu$^{2+}$ local spins become \emph{ferromagnetically} ordered at $T_\mathrm{m}\sim$ 18 K~\cite{rxs2014,nd2014,jin-Co,jin-Ir,adroja-Ir}. It has been concluded that SC appears only when $T_\mathrm{sc} > T_\mathrm{m}$ in systems of EuFe$_2$(As$_{1-x}$P$_x$)$_2$~\cite{cgh2011,jeevan}, Eu(Fe$_{1-x}$Co$_x$)$_2$As$_2$~\cite{nicklas}, and Sr$_{1-y}$Eu$_y$(Fe$_{0.88}$Co$_{0.12}$)$_2$As$_2$~\cite{SrEu122}. The conclusion also fits with the absence of SC in Eu(Fe$_{1-x}$Ni$_x$)$_2$As$_2$~\cite{ren2009Ni}, since the Eu-free analogous system Sr(Fe$_{1-x}$Ni$_x$)$_2$As$_2$ shows a maximum $T_\mathrm{sc}$ of 9.8 K, which is significantly lower than the expected $T_\mathrm{m}$~\cite{Sr122Ni}.

Very recently, a variant of EuFe$_2$As$_2$, i.e. the 1144-type $A$EuFe$_4$As$_4$ ($A$ = Rb and Cs), were synthesized and characterized~\cite{Eu1144,Rb1144.ly,Cs1144.ly}. The twin compounds adopt a crystal structure identical to that previously designed~\cite{1144.jh}, which was first realized in $AeA$Fe$_4$As$_4$ ($\emph{Ae}$ = Ca, Sr; $A$ = K, Rb, Cs)~\cite{1144}. In RbEuFe$_4$As$_4$, the Rb$^{+}$ and Eu$^{2+}$ planes, sandwiched by FeAs layers, stack alternately along the $c$ axis. The structure can also be viewed as an intergrowth between non-doped EuFe$_2$As$_2$ and heavily over-doped RbFe$_2$As$_2$. As a result, RbEuFe$_4$As$_4$ is intrinsically hole-doped, exhibiting SC at $T_{\mathrm{sc}}=$ 36.5 K without any SDW ordering~\cite{Eu1144,Rb1144.ly}. Additionally, evidence of FM of the Eu$^{2+}$ spins below $T_{\mathrm{m}}=$ 15 K is given by magnetization measurements~\cite{Rb1144.ly}. Compared with 122-type FSCs, the $T_{\mathrm{sc}}$ value is significantly higher, and the $T_{\mathrm{m}}$ value is slightly lower. Important to be noted is that the $T_{\mathrm{sc}}$ value of RbEuFe$_4$As$_4$ are almost the same as, or even larger than, those of the non-magnetic analogues (e.g. $T_{\mathrm{sc}}=$ 35.1 K in RbSrFe$_4$As$_4$~\cite{1144}), indicating that the Eu$^{2+}$ spins hardly break superconducting Cooper pairs. In this context, SC may survive easily in the presence of Eu-spin order, and therefore, it is of interest to seek for an SFM in the 1144-type system.

Now that RbEuFe$_4$As$_4$ bears an intrinsic hole doping (0.25 holes per Fe atom), it is natural to tune the $T_\mathrm{sc}$ value by hole depletion via electron doping. Our preliminary trial with a Ba-for-Rb substitution was demonstrated unsuccessful, since a 122-type phase, instead of the expected 1144-type phase, became stabilized. Then, we turned to a substitution at the Fe site. To compensate the doped holes more effectively, we chose Ni as the dopant, because Ni$^{2+}$ (3d$^{8}$) has two more itinerant electrons than Fe$^{2+}$ (3d$^{6}$) does, and more importantly, previous Ni-doping studies indeed show such an effect of electron doping~\cite{cgh2009,llj,Sr122Ni}.

In this paper, we report a systematic investigation on the magnetic and superconducting properties in RbEu(Fe$_{1-x}$Ni$_x$)$_4$As$_4$. As expected, $T_{\mathrm{sc}}$ decreases rapidly with the Ni doping. On the other hand, the Eu-spin FM and its $T_{\mathrm{m}}$ value remain unchanged. This leads to a discovery of SFMs in RbEu(Fe$_{1-x}$Ni$_x$)$_4$As$_4$ with $0.07\leq x\leq0.08$, in which SC survives when $T_{\mathrm{sc}}$ becomes lower than $T_{\mathrm{m}}$. The SFMs are found to show no Meissner state with a broadened resistive transition, because they are always under the internal field generated by the FM of Eu$^{2+}$ spins. The superconducting and magnetic phase diagram has been established, and the reason for the existence of SFMs is discussed.

\section{\label{sec:level2}Experimental Methods}

Polycrystalline samples of RbEu(Fe$_{1-x}$Ni$_x$)$_4$As$_4$ with $0 \leq x \leq 0.125$ were prepared by solid-state reactions in evacuated quartz ampoules sealed, similar to our previous report~\cite{Rb1144.ly}. The source materials were the constituent elements: Rb (99.75\%), As (99.999\%), Eu (99.9\%), Fe (99.998\%), and Ni (99.99\%), all from Alfa Aesar. Firstly, precursors of EuAs, FeAs, NiAs, and RbFe$_2$As$_2$ (with 5\% excess of Rb) were prepared by solid-state reactions in evacuated quartz tubes at 873-1023 K for 24 hours. These precursors and additional Fe powders were then mixed together in the nominal composition of RbEu(Fe$_{1-x}$Ni$_x$)$_4$As$_4$, followed by thoroughly homogenizing with ball milling in an Ar-filled glove box. Secondly, the mixtures were pressed into pellets which were loaded in an alumina container jacketed with a double-layer protector (a sealed Ta tube inside and a quartz ampoule outside). Finally, the samples were rapidly heated to 1133 K, holding for 20 hours, ended with quenching in cool water. To improve samples' purity, the synthesis was repeated once or twice, with an intermediate grinding.

Powder x-ray diffraction (XRD) was performed on a PANalytical x-ray diffractometer with a monochromatic Cu-K$_{\alpha1}$  radiation at room temperature. The lattice parameters were obtained by a least-squares fit of 15-25 reflections in the range of $5^{\circ}\leq 2\theta \leq 80^{\circ}$. The sample's chemical composition was checked by energy-dispersive x-ray spectroscopy (EDS). The resistivity and specific-heat measurements were conducted on a Quantum Design Physical Property Measurement System (PPMS-9). The dc magnetization was measured on a Quantum Design Magnetic Property Measurement System (MPMS-XL5).

\section{\label{sec:level3}Results and discussion}

\subsection{\label{subsec:level1}X-ray diffraction}

\begin{figure}
\includegraphics[width=8cm]{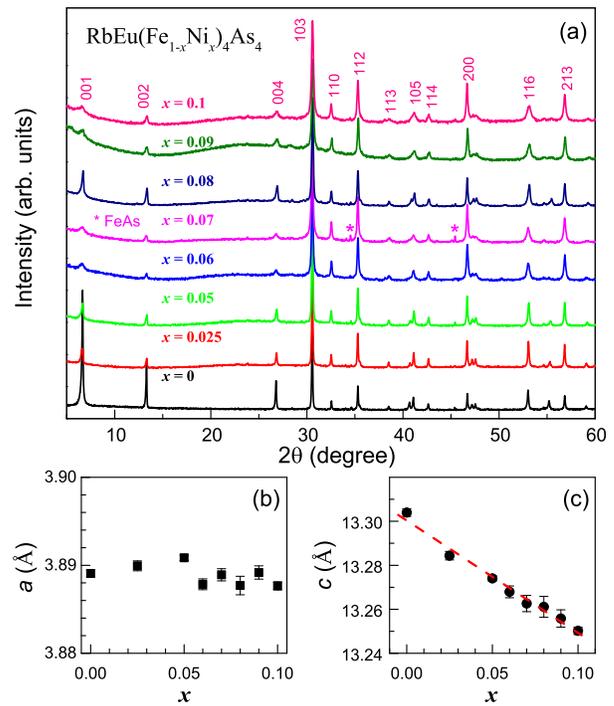}
\caption{(a) Powder X-ray diffraction patterns of RbEu(Fe$_{1-x}$Ni$_x$)$_4$As$_4$ ($0\leq x\leq0.1$) at room temperature. (b) and (c)  Lattice parameters $a$ and $c$ as functions of the nominal Ni concentration $x$. The dashed line in (c) gives the linear fit.}
\label{xrd}
\end{figure}

Figure~\ref{xrd}(a) shows XRD patterns of the series samples of RbEu(Fe$_{1-x}$Ni$_x$)$_4$As$_4$ ($0\leq x\leq0.1$), which can be well indexed by a tetragonal unit cell with $a \approx$ 3.89 {\AA} and $c \approx$ 13.3 {\AA}. The  variations in relative intensity is mainly due to (00l) preferred orientations. One sees that all the samples from $x$ = 0 to 0.1 are nearly single phase (only small amount of impurities such as FeAs appear in some of the samples). In the case of $x=0.125$, the XRD pattern (not shown) indicates formation of 122-type phase. Therefore, the present study is limited to samples with $0\leq x \leq0.1$.

\begin{table*}
\caption{Room-temperature lattice constants and physical-property parameters of RbEu(Fe$_{1-x}$Ni$_x$)$_4$As$_4$ ($0\leq x\leq0.1$). The digits in parentheses give twice of the standard deviation of the least-squares fit. $T_{\mathrm{sc}}^{\rho}$ is the superconducting midpoint transition temperature in the resistivity measurement. $T_{\mathrm{SDW}}$ denotes the spin-density-wave transition temperature. $T_{\mathrm{m}}$ and $\Theta$ are the magnetic-transition and Curie-Weiss temperatures, respectively. $\mu_{\mathrm{eff}}$ and $M_{\mathrm{sat}}$ are the effective magnetic moment in the paramagnetic state and the ordered moment in the ferromagnetic state, respectively. $H_{\mathrm{coe}}$ refers to the apparent coercive field.}
\begin{ruledtabular}
\begin{tabular}{cccccccccc}
 &\multicolumn{2}{c}{Lattice Constants}&\multicolumn{7}{c}{Physical-Property Parameters}\\

$x$ & $a$(\AA)&$c$ (\AA)&$T_{\mathrm{sc}}^{\rho}$ (K) & $T_{\mathrm{SDW}}$ (K)  & $T_{\mathrm{m}}$ (K) & $\Theta$ (K) & $\mu_{\mathrm{eff}}$ ($\mu_\mathrm{B}$/Eu)&$M_{\mathrm{sat}}$ ($\mu_\mathrm{B}$/Eu)&$H_{\mathrm{coe}}$ (Oe)\\
\hline
0  &3.8891(4)&13.304(17)&36.4 &- & 15.0 &23.6 &	7.95 &6.5& 360\\
0.025 &3.8900(6)&13.284(19)&30.3 &-& 15.0&24.3 &	7.88 &6.5&258\\
0.05 &3.8909(3)&13.274(12)&	23.0 &28.9& 15.0&24.5 &	8.00 &6.4&88\\
0.06 &3.8878(6)&13.268(27)&	18.1 &31.1& 15.0&24.3  & 7.65 &5.9&67\\
0.07 &3.8889(7)&13.263(37)&	11.2 &35.0& 15.1&24.2  & 7.46 &5.9&44\\
0.08 &3.8877(11)&13.261(48)&	2.1 &33.6& 14.7&24.4  &	7.85 &6.5&21\\
0.09 &3.8892(8)&13.256(40)&- &	31.3& 14.7&24.8 &7.79 &6.0&20\\
0.1 &3.8877(4)&13.250(17)& - & 29.4& 14.8&24.4 &	7.74 &6.3&24\\
\end{tabular}
\end{ruledtabular}
\label{Tab}
\end{table*}

The lattice constants were determined by a least-squares fit, the results of which are displayed in Table~\ref{Tab}. Figs.~\ref{xrd}(b,c) plot the fitted lattice parameters $a$ and $c$, respectively, as a function of the nominal Ni content. As is shown, while the $a$ axis of the unit cell basically remains unchanged, the $c$ axis decreases significantly with the Ni doping. The result is quite similar to those in Eu(Fe$_{1-x}$Ni$_x$)$_2$As$_2$~\cite{ren2009Ni}, LaFe$_{1-x}$Ni$_x$AsO~\cite{cgh2009}, and Ba(Fe$_{1-x}$Ni$_x$)$_2$As$_2$~\cite{llj} systems. The linear decrease in $c$, which obeys the Vegard's law, confirms that the dopant Ni indeed substitutes for Fe. The EDS on the sample of $x$ = 0.1 shows that the Ni content is 0.089(8), which remains unchanged throughout the sample. This confirms that the sample is homogeneous for the Ni doping, and the actual Ni-doping level is close to the nominal one within the measurement uncertainty.

\subsection{\label{subsec:level2}Magnetic properties}

\begin{figure}
\includegraphics[width=8cm]{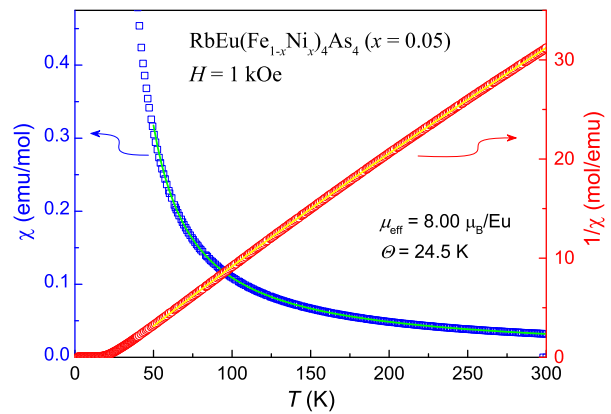}
\caption{Temperature dependence of dc magnetic susceptibility ($\chi=M/H$) for a typical sample of RbEu(Fe$_{1-x}$Ni$_x$)$_4$As$_4$ with $x$ = 0.05 at a magnetic field of $H$ = 1 kOe. The right-hand axis plots $1/\chi$, indicating dominant Curie-Weiss paramagnetism above 25 K. The data in the range of 50 K $\leq T \leq$ 300 K are fitted with Curie-Weiss law (displayed with solid lines), from which the effective local moment $\mu_{\mathrm{eff}}$ and the Curie-Weiss temperature $\Theta$ are extracted as shown. See the text for details.}
\label{cw}
\end{figure}

We first address the Eu-spin state by focusing on the high-temperature magnetic properties in RbEu(Fe$_{1-x}$Ni$_x$)$_4$As$_4$. As shown in Fig.~\ref{cw}, the magnetic susceptibility ($\chi$) at high temperatures is of Curie-Weiss-type paramagnetism. The $\chi(T)$ data in the temperature range of 50 K $\leq T \leq$ 300 K can be well fitted with an extended Curie-Weiss law, $\chi=\chi_0 + C/(T-\Theta)$, where $\chi_0$ is the temperature-independent term, $C$ gives the Curie constant from which the effective local moment is derived, and $\Theta$ represents the Curie-Weiss temperature. The derived effective moment $\mu_{\mathrm{eff}}$ (see Table~\ref{Tab}) ranges between 7.46 and 8.00 $\mu_\mathrm{B}$/f.u. (f.u. refers to formula unit), independent of the Ni substitution. These $\mu_{\mathrm{eff}}$ values are close to that (7.9 $\mu_\mathrm{B}$) expected for a free Eu$^{2+}$ ion, indicating a spin state with total spins of $S=7/2$ for the Eu$^{2+}$ ions. The $\Theta$ values fitted (from 23.6 to 24.8 K) is also independent of the Ni doping. The positive sign of $\Theta$ reflects dominant ferromagnetic interactions between Eu$^{2+}$ spins.

At low temperatures, the system undergoes superconducting and/or magnetic transitions. To determine their transition temperatures, $T_\mathrm{sc}$ and $T_\mathrm{m}$, we performed measurement of temperature-dependent magnetization, $M(T)$, under a low field of 10 Oe. In general, a superconducting transition can be easily recognized by the strong diamagnetic signal owing to Meissner effect. However, in cases of the coexistence between SC and FM with $T_\mathrm{sc} \leq T_\mathrm{m}$, the diamagnetic signal may be covered up by FM. For the ferromagnetic transition, on the other hand, the Curie temperature is traditionally determined by Arrot approach~\cite{Arrot}. However, the presence of SC makes the method invalid. Here we take advantage of the magnetic hysteresis arising from the appearance of magnetic domains. Namely, $T_\mathrm{m}$ is defined by the bifurcation temperature between field-cooling (FC) and zero-field-cooling (ZFC) $M(T)$ data measured under magnetic fields lower than the coercive field. In fact, the bifurcation point basically coincides with the kink (peak) in the FC (ZFC) curves. Furthermore, the resultant $T_\mathrm{m}$ value is precisely consistent with the heat-capacity measurement~\cite{Rb1144.ly}. Note that a type-II SC may also give rise to a bifurcation between FC and ZFC curves owing to magnetic flux-pinning effect.

\begin{figure*}
\includegraphics[width=17cm]{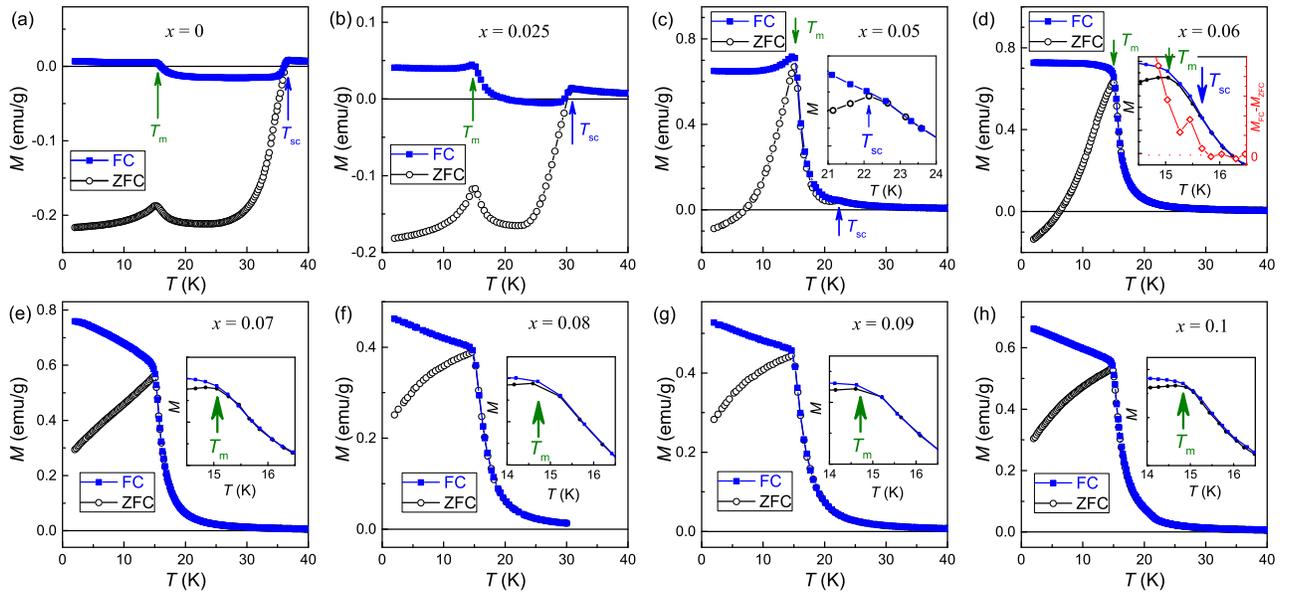}
\caption{Temperature dependence of magnetization under a magnetic field of 10 Oe for RbEu(Fe$_{1-x}$Ni$_x$)$_4$As$_4$ ($0\leq x\leq0.1$) samples. FC (solid symbols) and ZFC (open symbols) denote field cooling and zero-field cooling, respectively, in the magnetic measurements. $T_\mathrm{sc}$ and $T_\mathrm{m}$ marked with arrows represent the superconducting and magnetic transition temperatures, respectively. The insets show close-ups near the superconducting/magnetic transitions. The inset in panel (d) also plots the difference between the FC and ZFC data, using the righ-hand axis.}
\label{lfmt}
\end{figure*}

Figure~\ref{lfmt}(a-h) shows the low-field $M(T)$ data for samples of the RbEu(Fe$_{1-x}$Ni$_x$)$_4$As$_4$ series. The non-doped ($x$ = 0) compound shows SC at $T_\mathrm{sc}$ = 36.5 K and FM below $T_\mathrm{m}$ = 15 K~\cite{Rb1144.ly}. For $x$ = 0.025, the superconducting transition keeps robust, yet with an obviously reduced $T_\mathrm{sc}$ of 30.5 K. At $x$ = 0.05, in which $T_\mathrm{sc}$ is further decreased to 22 K, the superconducting transition becomes much less remarkable. For $x$ = 0.06, only weak signature of SC at 15.7 K can be traced from the slight difference between the FC and ZFC data, the latter of which comes from flux-pinning effect. The existence of SC is also evidenced from the diamagnetism in the ZFC data at lower temperatures and, in particular, from the zero resistance at 16 K in the resistivity measurement shown below. In the cases of $x\geq$ 0.07, no magnetic signal for SC can be detected, although the resistivity measurement clearly shows superconducting transitions at lower temperatures for $0.07\leq x\leq0.08$.

In contrast with the monotonic suppression in $T_\mathrm{sc}$ with increasing Ni doping, the magnetic transition temperature $T_\mathrm{m}$ almost keeps unchanged. Note that the Curie-Weiss temperature $\Theta$, which is remarkably higher than $T_\mathrm{m}$, does not depend on the Ni doping either. The lower-than-expected $T_\mathrm{m}$ value is probably related to the quasi-two-dimensional magnetism caused by a much weaker magnetic coupling along the $c$ axis~\cite{Rb1144.ly}. We will discuss on the magnetic interactions later on.

\begin{figure}
\includegraphics[width=8.5cm]{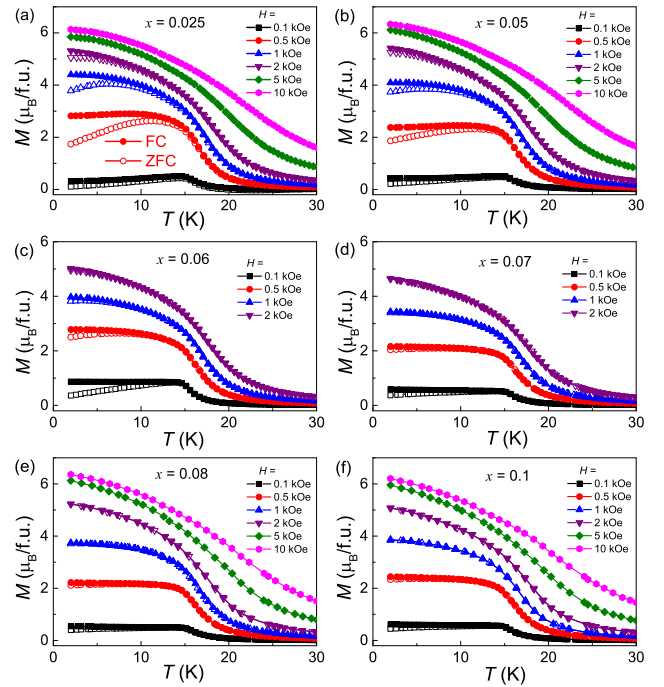}
 \caption{Temperature dependence of field-cooling (FC, solid symbols) and zero-field-cooling (ZFC, open symbols) magnetization under different magnetic fields for $x$ = 0.025 (a), 0.05 (b), 0.06 (c), 0.07 (d), 0.08 (e), and 0.1 (f) in RbEu(Fe$_{1-x}$Ni$_x$)$_4$As$_4$. The magnetization is conversed into Bohr magnetons per formula unit.}
 \label{hfmt}
\end{figure}

One may also note that, for lower Ni doping with $x\leq$ 0.05, the FC magnetization \emph{decreases} with decreasing temperature just below $T_\mathrm{m}$, showing a peak-like anomaly (PLA) at $T_\mathrm{m}$, which casts doubt on the nature of the magnetic transition. To address this issue, we measured the $M(T)$ data at elevated magnetic fields shown in Fig.~\ref{hfmt}. One sees that the PLA disappears under relatively low fields ($\sim$0.5 kOe). At higher fields with $H\geq$ 5 kOe, all the $M(T)$ curves become almost identical without bifurcations in the FC and ZFC data, suggesting commonality of FM in the system. In fact, a similar PLA behavior is also seen in 122-type FSCs~\cite{jiang2009,zapf2013} where an Eu-spin FM is unambiguously demonstrated~\cite{rxs2014,nd2014,jin-Co}. Note that the PLA here happens only for $T_\mathrm{sc}>T_\mathrm{m}$. This suggests that the PLA is probably in relation with the presence of SC. As is pointed out theoretically, the state of FM may be modified into crypto-ferromagnetism~\cite{Anderson} or dense-domain structure~\cite{Buzdin} owing to the presence of SC. The expected fine ferromagnetic domains that are antiferromagnetically aligned could give rise to a PLA in the FC $M(T)$ curve. Future studies with single-crystalline samples by magnetic-force microscopy~\cite{Eu122P.MFM} seem promising to clarify this issue.

\begin{figure}
\includegraphics[width=8.5cm]{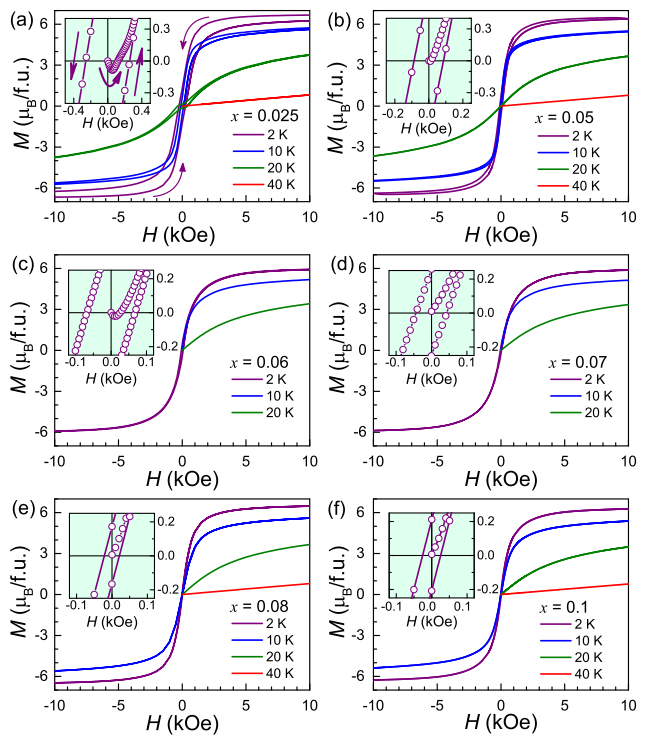}
 \caption{Isothermal magnetization at low temperatures in the RbEu(Fe$_{1-x}$Ni$_x$)$_4$As$_4$ series. The insets show close-ups of the magnetic hysteresis  at 2 K, from which the virgin magnetization as well as the coercive field can be seen clearly.}
 \label{mh}
\end{figure}

The Eu-spin FM in RbEu(Fe$_{1-x}$Ni$_x$)$_4$As$_4$ is further confirmed by the isothermal magnetization, $M(H)$, shown in Fig.~\ref{mh}(a-f). At high temperatures (e.g., 40 K), the $M(H)$ data are essentially linear, consistent with the paramagnetic state of Eu$^{2+}$ spins. At temperatures below $T_\mathrm{m}$, by contrast, the $M(H)$ curves are of an S shape, characteristic of an FM. The saturation magnetization $M_{\mathrm{sat}}$, defined here as the magnetization value at 1 T and 2 K, scatters from  5.9 to 6.5 $\mu_\mathrm{B}$/f.u. (see Table~\ref{Tab}). Samples containing more FeAs impurity tend to have relatively low $M_{\mathrm{sat}}$ (and $\mu_{\mathrm{eff}}$ also). Other samples show an $M_{\mathrm{sat}}$ value that is close to the expected one (7.0 $\mu_\mathrm{B}$ per Eu$^{2+}$)~\cite{note2}, indicating that Eu$^{2+}$ spins order ferromagnetically.

Note that for samples of $x\leq$ 0.05, which have higher $T_{\mathrm{sc}}$ values, the magnetic hysteresis is extended to high fields where the magnetization is saturated. This clearly indicates the existence of type-II SC that commonly exhibits flux-pinning effect, the latter of which also gives rise to an enhancement of the apparent coercive field $H_{\mathrm{coe}}$. As shown in Table~\ref{Tab}, the intrinsic $H_{\mathrm{coe}}$, given by non-superconducting samples of $0.08< x\leq0.1$, is actually around 20 Oe. The sample of $x=0.07$ shows an enhanced $H_{\mathrm{coe}}$ of 44 Oe, implying the existence of SC. Indeed, the resistivity measurement below demonstrates a superconducting transition at $T_{\mathrm{sc}}=$ 11 K, 4 K lower than $T_{\mathrm{m}}$. That is to say, the sample is actually an SFM. Notably, no superconducting diamagnetism is detected by the $M(T)$ data [Fig.~\ref{lfmt}(e)] and the virgin $M(H)$ curve [Fig.~\ref{mh}(c)]. This observation is consistent with the absence of Meissner state, as expected from the internal field ($\sim$4.5 kOe~\cite{note3}) that is much higher than the intrinsic lower critical field. It is of great interest for the future to look into the anisotropic magnetic properties with using the single-crystalline samples.

\subsection{\label{subsec:level3}Electrical resistivity}

\begin{figure}
\includegraphics[width=8cm]{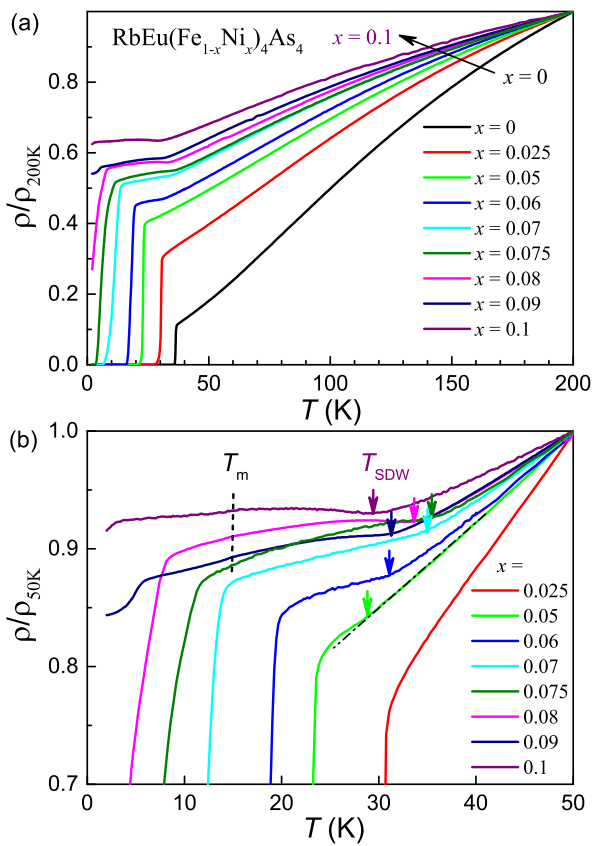}
 \caption{Temperature dependence of resistivity for RbEu(Fe$_{1-x}$Ni$_x$)$_4$As$_4$ polycrystalline samples at zero field. The top and bottom panels adopt a normalized scale relative to the resistivity values at 200 and 50 K, respectively. In the bottom panel, the arrows labeled with $T_{\mathrm{SDW}}$ point to the resistivity upturn that probably comes from a spin-density wave (SDW) transition. The left dashed line labeled with $T_{\mathrm{m}}$ marks the tiny kinks at which the Eu$^{2+}$ spins order ferromagnetically.}
 \label{rt}
\end{figure}

Figure~\ref{rt} shows the electrical resistivity ($\rho$) for RbEu(Fe$_{1-x}$Ni$_x$)$_4$As$_4$ polycrystalline samples. To highlight the evolution of the temperature dependence, and also to present the superconducting/SDW transitions clearly, we normalize the $\rho(T)$ data relative to the resistivity values at 200 and 50 K, respectively. First of all, the slope d$\rho$/d$T$ in the normal state decreases monotonically with the Ni doping, giving rise to an increase in the residual resistivity at low temperatures. This is consistent with the increase of Fe-site disorder with Ni doping. Secondly, the superconducting transition temperature $T_{\mathrm{sc}}^{\rho}$ decreases with the Ni doping, and SC is completely suppressed at $x=0.09$ (the slight drop around 5 K is probably due to sample's inhomogeneity). The result is basically consistent with the magnetic measurement for $x\geq0.06$ with $T_{\mathrm{sc}}> T_{\mathrm{m}}$. In the case of $T_{\mathrm{sc}}\leq T_{\mathrm{m}}$, nevertheless, SC cannot be directly detected in the magnetic measurement above, and the resistive transitions become remarkably broadened. The broadened resistive transition is similar to the observation in Eu(Fe$_{0.81}$Co$_{0.19}$)$_2$As$_2$~\cite{Eu122Co.Tran} also with $T_{\mathrm{sc}}< T_{\mathrm{m}}$, which can be explained in terms of the dissipative flow of spontaneous vortices~\cite{SVP.jiao}.

Another interestingly point is that the samples with $x\geq$ 0.05 show a resistivity upturn above $T_{\mathrm{sc}}$, which is probably due to a spin-density-wave (SDW) ordering in the Fe sublattice. At $x=$ 0.05, the hole concentration is reduced to $n_\mathrm{h}$ = 0.15 holes per Fe atom, if assuming that every doped Ni atom depletes two holes. In the prototype Ba$_{1-x}$K$_x$Fe$_2$As$_2$ system, SDW order remains at $n_\mathrm{h}$ = 0.15~\cite{BaK122.cxh}. Note that the SDW transition temperature $T_{\mathrm{SDW}}$ here is much lower than expected. Furthermore, $T_{\mathrm{SDW}}$ decreases with $x$ in the high doping regime. These results can be ascribed to the Fe-site disorder mentioned above. Similarly, a recovery of SDW by charge compensation was reported in Ba$_{1-x}$K$_x$Fe$_{1.86}$Co$_{0.14}$As$_2$~\cite{BaK122Co} and  Eu$_{0.5}$K$_{0.5}$(Fe$_{1-x}$Ni$_x$)$_2$As$_2$~\cite{EuK122Ni} systems. Here we note that, according to a recent report on Ni- and
Co-doped CaKFe$_4$As$_4$~\cite{1144Ni.canfield}, the recovered SDW phase may have strikingly different magnetic order.

In addition to the SDW-like anomaly above, Fig.~\ref{rt}(b) also shows a very slight (yet observable) kink at $T_{\mathrm{m}}$. This is due to the reduction of magnetic scattering on the charge carriers when Eu$^{2+}$ spins become ordered, akin to the case in EuFe$_2$As$_2$~\cite{ren2008}. The tiny resistivity change at $T_{\mathrm{m}}$ seems to reflect weak interactions between Eu$^{2+}$ spins and the charge carriers.

\subsection{\label{subsec:level3}Heat capacity}

\begin{figure}
\includegraphics[width=8cm]{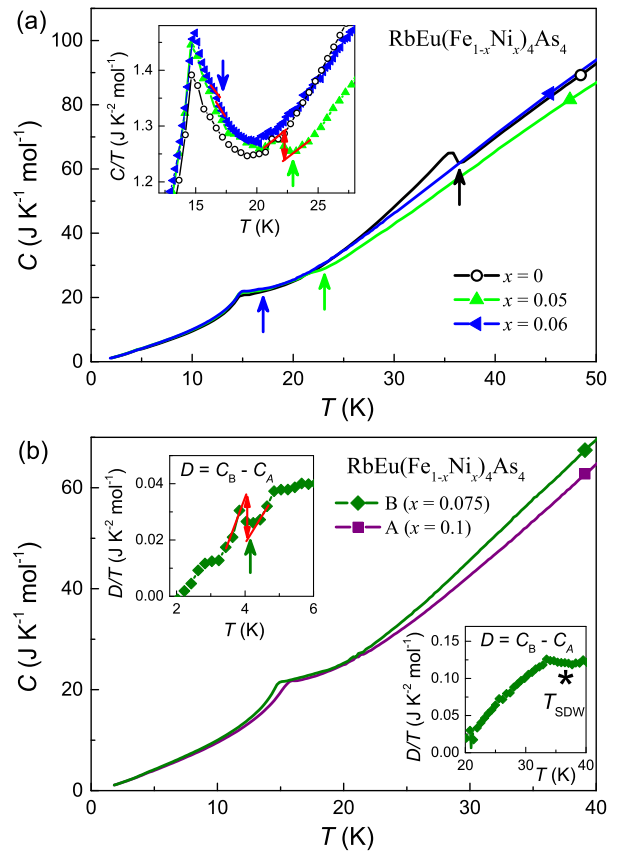}
 \caption{Temperature dependence of specific heat capacity ($C$) of the ferromagnetic superconductors (a) and the superconducting ferromagnet (b) in RbEu(Fe$_{1-x}$Ni$_x$)$_4$As$_4$ under zero magnetic field. The superconducting transitions are marked by arrows. Note that the insets in top and bottom panels plot $C/T$ and $D/T$, respectively, where $D$ stands for the specific-heat difference between samples of $x=$ 0.075 and 0.1.}
 \label{hc}
\end{figure}

Figure~\ref{hc} shows the temperature dependence of specific heat of RbEu(Fe$_{1-x}$Ni$_x$)$_4$As$_4$ under zero field, tracking for superconducting transitions in the FSCs and SFMs. For $x$ = 0, the specific-heat jump is clearly seen with $\Delta C/T_{\mathrm{sc}}$ = 208 mJ K$^{-2}$ mol$^{-1}$ at the superconducting transition~\cite{Rb1144.ly}. With increasing the Ni doping, the specific-heat jump becomes unapparent. At $x$ = 0.05 and 0.06, for example, the $\Delta C/T_{\mathrm{sc}}$ values are estimated to be $\sim$50 and $\sim$20 mJ K$^{-2}$ mol$^{-1}$, respectively. Similar dramatic reduction in $\Delta C$ was also observed in underdoped Ba$_{1-x}$K$_x$Fe$_2$As$_2$~\cite{BaK122.HC}. As a comparison, the underdoped Ba$_{0.77}$K$_{0.23}$Fe$_2$As$_2$ sample ($T_{\mathrm{sc}}$ = 23 K) shows a $\Delta C/T_{\mathrm{sc}}$ value of $\sim$40 mJ K$^{-2}$ mol$^{-1}$~\cite{BaK122.HC}, comparable to that of the $x$ = 0.05 sample (with a similar $T_{\mathrm{sc}}$) in the present system. Note that the strength of superconducting coupling varies with the doping level~\cite{BaK122.HC}, the former of which is reflected by the dimensionless parameter, $p=\Delta C/\gamma_\mathrm{n}T_{\mathrm{sc}}$, where $\gamma_\mathrm{n}$ denotes the Sommerfeld coefficient in the normal state. The drastic reduction in $\Delta C$ for the underdoped samples is actually a consequence of the \emph{concurrent} decrease in all the three factors: $\gamma_\mathrm{n}$, $T_{\mathrm{sc}}$ and $p$.

In the case of SFMs with $T_{\mathrm{sc}}<T_{\mathrm{m}}$, no specific-heat anomaly at $T_{\mathrm{sc}}$ can be directly seen from the raw data. Nevertheless, by subtraction of the specific heat between the superconducting ($x$ = 0.075) and non-superconducting ($x$ = 0.1) samples, the specific-heat jump is still observable at 4.1 K (at which the resistivity drop to 6\%), as shown in the upper-left inset of Fig.~\ref{hc}(b) (note that the kink at $\sim$5 K is probably due to an antiferromagnetic transition from the very small amount of Eu$_3$O$_4$ impurity~\cite{Eu3O4}). Since the Sommerfeld coefficient is greatly reduced in the underdoped regime~\cite{BaK122.HC}, the $\Delta C/T_{\mathrm{sc}}$ value ($\sim$10 mJ K$^{-2}$ mol$^{-1}$) is actually appreciable, which supports the bulk nature of superconductivity. Additionally, from the specific-heat difference shown in the lower inset, one can see another anomaly at $\sim$36 K (albeit no anomaly is observable again in the raw data), in accordance with the resistivity upturn. This anomaly, if being intrinsic, should be related to the recovered SDW transition.

\subsection{\label{subsec:level4}Phase diagram}

\begin{figure}
\includegraphics[width=8cm]{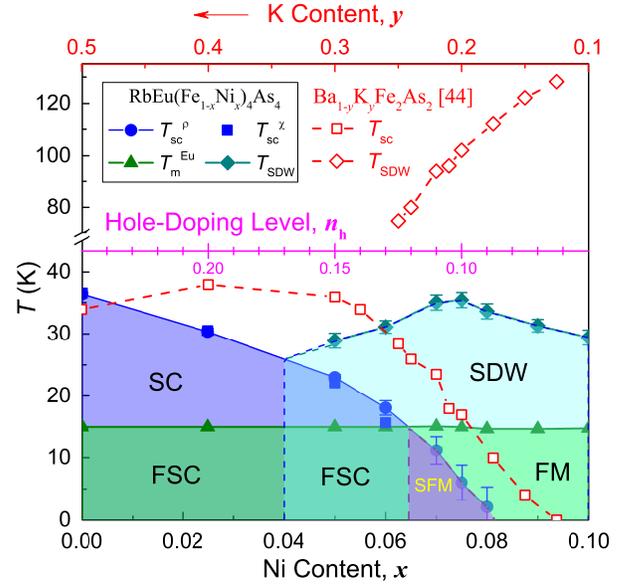}
 \caption{Superconducting and magnetic phase diagram in RbEu(Fe$_{1-x}$Ni$_x$)$_4$As$_4$ containing various electronic phases highlighted with colors. $T_{\mathrm{sc}}^{\rho}$ is the midpoint temperature of superconducting resistive transitions, and the error bars denote the transition widths. $T_{\mathrm{sc}}^{\chi}$ is the onset temperature of superconducting diamagnetic transitions. Other abbreviations include: SC, superconductivity; SDW, spin-density wave; FM, ferromagnet; FSC, ferromagnetic superconductor; SFM, superconducting ferromagnet. For comparison, the phase lines of Ba$_{1-y}$K$_y$Fe$_2$As$_2$~\cite{BaK122.Kanatzidis} are plotted using the top axis. Both horizontal axes share the same hole-doping level, $n_\mathrm{h}=0.25-2x=y/2$, as shown in the middle axis.}
 \label{pd}
\end{figure}

The results above allow us to construct the phase diagram in RbEu(Fe$_{1-x}$Ni$_x$)$_4$As$_4$, which is displayed in Fig.~\ref{pd}. As usual, the bottom axis employs the direct control parameter, i.e. the Ni content $x$. Since substitution of Fe$^{2+}$ (3d$^{6}$) with Ni$^{2+}$ (3d$^{8}$) doubly compensates the self-doped holes, the expected hole concentration is $n_\mathrm{h} = 0.25-2x$, which is also shown in the middle horizontal axis. With the hole depletion by Ni doping, $T_{\mathrm{sc}}$ decreases monotonically. Note that $T_{\mathrm{sc}}$ decreases more rapidly for $x\geq0.05$ where SDW order appears, suggesting competing nature between SC and SDW. SC disappears at $x>0.08$ or at $n_\mathrm{h}\leq0.09$. One notes that $T_{\mathrm{SDW}}$ goes down for $x\geq0.075$.

To understand the possible role of disorder, we compare with the electronic phase diagram of Ba$_{1-y}$K$_y$Fe$_2$As$_2$~\cite{BaK122.Kanatzidis} which is disorder-free at the Fe site. One sees that the $T_{\mathrm{sc}}$ values in RbEu(Fe$_{1-x}$Ni$_x$)$_4$As$_4$ are overall lower in the doping area, and the critical hole concentration (for appearance of SC) is significantly higher. On the other hand, the $T_{\mathrm{SDW}}$ values are even much lower, especially in the high doping regime. Both results strongly suggest that the disorder by Ni doping plays an important role in suppressing SC as well as SDW.

Then, how much do the Eu$^{2+}$ spins influence the $T_{\mathrm{sc}}$? As we emphasize in the Introduction, first of all, SC in RbEuFe$_4$As$_4$ is not suppressed at all. Secondly, the $T_{\mathrm{sc}}$ values in RbEu(Fe$_{1-x}$Ni$_x$)$_4$As$_4$ are on average $\sim$10 K lower than those in Ba$_{1-y}$K$_y$Fe$_2$As$_2$~\cite{BaK122.Kanatzidis}. The amount of $T_{\mathrm{sc}}$ reduction is very close to that of Eu-free Ba$_{1-y}$K$_y$Fe$_{1.86}$Co$_{0.14}$As$_2$ (compared in the same way with Ba$_{1-y}$K$_y$Fe$_2$As$_2$)~\cite{BaK122Co}, suggesting that disorder plays the dominant role for suppression of $T_{\mathrm{sc}}$. In other words, the Eu$^{2+}$ spins play a relatively minor (if not none at all) role in suppressing SC. Thirdly, the $T_{\mathrm{sc}}$ values of RbEu(Fe$_{1-x}$Ni$_x$)$_4$As$_4$ are even higher than those in Eu-diluted Eu$_{0.5}$K$_{0.5}$(Fe$_{1-x}$Ni$_x$)$_2$As$_2$~\cite{EuK122Ni}, the latter of which shows an antiferromagnetism (e.g., $T_{\mathrm{N}}=$ 8.5 K for $x=0.08$) for Eu$^{2+}$ spins which implies relatively less influence on SC. This comparison further corroborates that Eu$^{2+}$ spins hardly suppress $T_{\mathrm{sc}}$ in the 1144-type system.

In contrast to the dramatic changes in the states associated with the Fe sublattice, the ferromagnetic state in the Eu sublattice remains unchanged. In particular, $T_{\mathrm{m}}$ does not depend on the Ni doping, resulting in a crossing at $x\sim0.065$ between the data lines of $T_{\mathrm{sc}}$ and $T_{\mathrm{m}}$. According to the classification for materials with coexistence of SC and FM~\cite{chucw,SVP.jiao}, the system changes from FSC to SFM at the crossing point. For $x>0.08$, the system shows coexistence of Eu-spin FM and Fe-site SDW below 15 K. Although the sample with $x=0.125$ (corresponding to $n_\mathrm{h}=0$) could not be synthesized owing to the solubility limit, one may expect by extrapolation that this completely hole-compensated material would show a similar behavior to that of the $x=0.1$ sample.

\subsection{\label{subsec:level5}Discussion}

In the following, we discuss why both FSCs and SFMs exist in RbEu(Fe$_{1-x}$Ni$_x$)$_4$As$_4$ system. First of all, coexistence of SC and FM in Eu-containing 122-type iron pnictides was tentatively explained in terms of Fe-3d multi-orbitals and robustness of SC~\cite{cgh2012}. On the one hand, multi-3d-orbitals in the valence band allow both SC mainly from 3d$_{yz/zx}$ electrons and Eu-spin FM via Ruderman-Kittel-Kasuya-Yosida (RKKY) interactions through 3d$_{x^{2}-y^{2}}$ and 3d$_{z^{2}}$ electrons. On the other hand, the intrinsic upper critical field $H_{\mathrm{c2}}^*$ of the superconductors alike (e.g., an Eu-free analogue) could be high enough to overcome the exchange field between Eu$^{2+}$ spins and Cooper pairs.

In this context, the absence of SC for $T_\mathrm{sc} < T_\mathrm{m}$ in EuFe$_2$(As$_{1-x}$P$_x$)$_2$~\cite{cgh2011,jeevan}, Eu(Fe$_{1-x}$Ni$_x$)$_2$As$_2$~\cite{ren2009Ni}, and Eu(Fe$_{1-x}$Co$_x$)$_2$As$_2$~\cite{nicklas} can be attributed to the relatively low $H_{\mathrm{c2}}^*$ in relation with the lower $T_\mathrm{sc}$. Here we note that the Eu(Fe$_{0.81}$Co$_{0.19}$)$_2$As$_2$ single crystals grown from Sn flux were reported to show SC with $T_\mathrm{sc} < T_\mathrm{m}$~\cite{Eu122Co.Tran}. The possible reason is that the Eu-spin exchange field could be reduced due to existence of defects in the Eu sublattice. One notes that the Sn-flux-grown ``Eu(Fe$_{0.82}$Co$_{0.18}$)$_2$As$_2$" samples showed SDW order~\cite{Eu122Co.jwt}, indicating that they were actually in an underdoped regime. The underdoped status with heavy Co-doping levels suggests the possibility of significant Eu deficiencies. Besides, the flux Sn could also be incorporated into the Eu site~\cite{Sn-flux}. Both factors lead to dilution in the Eu sublattice, such that the exchange field that breaks Cooper pairs may be reduced, which helps the survival of SC.

In the 1144-type system of RbEu(Fe$_{1-x}$Ni$_x$)$_4$As$_4$, the Eu$^{2+}$ spins hardly suppress SC. Therefore, not only do FSCs exist, but also SFMs occur in RbEu(Fe$_{1-x}$Ni$_x$)$_4$As$_4$. Note that the internal field generated from the Eu-spin FM is about 4.5 kOe, being high enough to induce spontaneous vortices, yet not high enough to destroy SC.


Finally, we comment on the magnetic interactions between Eu$^{2+}$ spins in RbEu(Fe$_{1-x}$Ni$_x$)$_4$As$_4$. The experimental fact is that neither $T_{\mathrm{m}}$ nor $\Theta$ change with the Ni doping. This result contrasts to the change in $T_{\mathrm{m}}$ from 20 to 16 K, accompanying with an antiferromagnetic-to-ferromagnetic transition, only by 3\% Ni doping in Eu(Fe$_{1-x}$Ni$_x$)$_2$As$_2$~\cite{ren2009Ni}. The sensitivity to Ni doping dictates an indirect RKKY interaction whose strength is proportional to $\mathrm{cos}(2k_{\mathrm{F}}r)/r^{3}$, where $k_\mathrm{F}$ is the Fermi vector and, $r$ is the distance between Eu$^{2+}$ spins. Conversely, the invariance of $T_{\mathrm{m}}$ and $\Theta$ against electron doping in RbEu(Fe$_{1-x}$Ni$_x$)$_4$As$_4$ suggests that the RKKY interaction may not be the dominant magnetic exchange interaction. This reminds us of the ferromagnetic europium chalcogenides, EuO ($T_\mathrm{m}=69.2$ K) and EuS ($T_\mathrm{m}=16.6$ K)~\cite{EuCh.neutron}, where there are no itinerant electrons for an indirect RKKY interaction. So, the effective ferromagnetic couplings between Eu$^{2+}$ spins in RbEu(Fe$_{1-x}$Ni$_x$)$_4$As$_4$ may be due to the so-called d$-$f~\cite{Kasuya} and/or As$-$Eu$-$As superexchange interactions. Such exchange interactions naturally explain the decoupling between Eu-4f and Fe-3d orbitals, which conversely sheds light on the mechanism of iron-based superconductivity. Future theoretical analyses and calculations may help to clarify this issue.

\section{\label{sec:level4}Conclusion}

In summary, we have systematically studied the magnetic and superconducting properties in RbEu(Fe$_{1-x}$Ni$_x$)$_4$As$_4$ ($0\leq x\leq0.1$). With the Ni doping that introduces extra itinerant electrons, the self-doped holes are gradually compensated. Resultantly, the superconducting transition temperature $T_{\mathrm{sc}}$ decreases rapidly, and superconductivity disappears at $x\sim0.08$. The hole depletion also brings the recovery of SDW order for $x\geq0.05$. For the Eu sublattice, the Eu-spin ferromagnetism in RbEuFe$_4$As$_4$ remains, and its Curie temperature keeps unchanged. This gives rise to unique SFMs showing absence of Meissner state. The realization of crossover from FSCs to SFMs makes RbEu(Fe$_{1-x}$Ni$_x$)$_4$As$_4$ a promising playground to look into the interplay between SC and FM for the future.

\

\begin{acknowledgments}
This work was supported by the National Natural Science Foundation of China (No. 11474252) and National Key Research and Development Program of China (Nos. 2016YFA0300202).
\end{acknowledgments}



\begin{thebibliography}{56}%
\makeatletter
\providecommand \@ifxundefined [1]{%
 \@ifx{#1\undefined}
}%
\providecommand \@ifnum [1]{%
 \ifnum #1\expandafter \@firstoftwo
 \else \expandafter \@secondoftwo
 \fi
}%
\providecommand \@ifx [1]{%
 \ifx #1\expandafter \@firstoftwo
 \else \expandafter \@secondoftwo
 \fi
}%
\providecommand \natexlab [1]{#1}%
\providecommand \enquote  [1]{``#1''}%
\providecommand \bibnamefont  [1]{#1}%
\providecommand \bibfnamefont [1]{#1}%
\providecommand \citenamefont [1]{#1}%
\providecommand \href@noop [0]{\@secondoftwo}%
\providecommand \href[0]{\begingroup \@sanitize@url \@href}%
\providecommand \@href[1]{\@@startlink{#1}\@@href}%
\providecommand \@@href[1]{\endgroup#1\@@endlink}%
\providecommand \@sanitize@url [0]{\catcode `\\12\catcode `\$12\catcode
  `\&12\catcode `\#12\catcode `\^12\catcode `\_12\catcode `\%12\relax}%
\providecommand \@@startlink[1]{}%
\providecommand \@@endlink[0]{}%
\providecommand \url  [0]{\begingroup\@sanitize@url \@url }%
\providecommand \@url [1]{\endgroup\@href {#1}{\urlprefix }}%
\providecommand \urlprefix  [0]{URL }%
\providecommand \Eprint [0]{\href }%
\providecommand \doibase [0]{http://dx.doi.org/}%
\providecommand \selectlanguage [0]{\@gobble}%
\providecommand \bibinfo  [0]{\@secondoftwo}%
\providecommand \bibfield  [0]{\@secondoftwo}%
\providecommand \translation [1]{[#1]}%
\providecommand \BibitemOpen [0]{}%
\providecommand \bibitemStop [0]{}%
\providecommand \bibitemNoStop [0]{.\EOS\space}%
\providecommand \EOS [0]{\spacefactor3000\relax}%
\providecommand \BibitemShut  [1]{\csname bibitem#1\endcsname}%
\let\auto@bib@innerbib\@empty
\bibitem [{\citenamefont {Bulaevskii}\ \emph {et~al.}(1985)\citenamefont
  {Bulaevskii}, \citenamefont {Buzdin}, \citenamefont {Kuli\'{c}},\ and\
  \citenamefont {Panjukov}}]{buzdin1985}%
  \BibitemOpen
  \bibfield  {author} {\bibinfo {author} {\bibfnamefont {L.~N.}\ \bibnamefont
  {Bulaevskii}}, \bibinfo {author} {\bibfnamefont {A.~I.}\ \bibnamefont
  {Buzdin}}, \bibinfo {author} {\bibfnamefont {M.~L.}\ \bibnamefont
  {Kuli\'{c}}}, \ and\ \bibinfo {author} {\bibfnamefont {S.~V.}\ \bibnamefont
  {Panjukov}},\ }\bibinfo {title} {Coexistence of superconductivity and
  magnetism theoretical predictions and experimental results},\ \href{\doibase
  10.1080/00018738500101741} {\bibfield  {journal} {\bibinfo  {journal} {Adv.
  Phys.}\ }\textbf {\bibinfo {volume} {34}},\ \bibinfo {pages} {175} (\bibinfo
  {year} {1985})}\BibitemShut {NoStop}%
\bibitem [{\citenamefont {Buzdin}(2005)}]{buzdin2005}%
  \BibitemOpen
  \bibfield  {author} {\bibinfo {author} {\bibfnamefont {A.~I.}\ \bibnamefont
  {Buzdin}},\ }\bibinfo {title} {Proximity effects in
  superconductor-ferromagnet heterostructures},\ \href{\doibase
  10.1103/RevModPhys.77.935} {\bibfield  {journal} {\bibinfo  {journal} {Rev.
  Mod. Phys.}\ }\textbf {\bibinfo {volume} {77}},\ \bibinfo {pages} {935}
  (\bibinfo {year} {2005})}\BibitemShut {NoStop}%
\bibitem [{\citenamefont {Cao}\ \emph {et~al.}(2012)\citenamefont {Cao},
  \citenamefont {Jiao}, \citenamefont {Luo}, \citenamefont {Ren}, \citenamefont
  {Jiang},\ and\ \citenamefont {Xu}}]{cgh2012}%
  \BibitemOpen
  \bibfield  {author} {\bibinfo {author} {\bibfnamefont {G.-H.}\ \bibnamefont
  {Cao}}, \bibinfo {author} {\bibfnamefont {W.-H.}\ \bibnamefont {Jiao}},
  \bibinfo {author} {\bibfnamefont {Y.-K.}\ \bibnamefont {Luo}}, \bibinfo
  {author} {\bibfnamefont {Z.}~\bibnamefont {Ren}}, \bibinfo {author}
  {\bibfnamefont {S.}~\bibnamefont {Jiang}}, \ and\ \bibinfo {author}
  {\bibfnamefont {Z.-A.}\ \bibnamefont {Xu}},\ }\bibinfo {title} {Coexistence
  of superconductivity and ferromagnetism in iron pnictides},\
  \href{http://stacks.iop.org/1742-6596/391/i=1/a=012123} {\bibfield  {journal}
  {\bibinfo  {journal} {J. Phys.: Conf. Ser.}\ }\textbf {\bibinfo {volume}
  {391}},\ \bibinfo {pages} {012123} (\bibinfo {year} {2012})}\BibitemShut
  {NoStop}%
\bibitem [{\citenamefont {Sonin}\ and\ \citenamefont {Felner}(1998)}]{Felner}%
  \BibitemOpen
  \bibfield  {author} {\bibinfo {author} {\bibfnamefont {E.~B.}\ \bibnamefont
  {Sonin}}\ and\ \bibinfo {author} {\bibfnamefont {I.}~\bibnamefont {Felner}},\
  }\bibinfo {title} {Spontaneous vortex phase in a superconducting weak
  ferromagnet},\ \href{\doibase 10.1103/PhysRevB.57.R14000} {\bibfield
  {journal} {\bibinfo  {journal} {Phys. Rev. B}\ }\textbf {\bibinfo {volume}
  {57}},\ \bibinfo {pages} {14000(R)} (\bibinfo {year} {1998})}\BibitemShut
  {NoStop}%
\bibitem [{\citenamefont {Lorenz}\ and\ \citenamefont {Chu}(2005)}]{chucw}%
  \BibitemOpen
  \bibfield  {author} {\bibinfo {author} {\bibfnamefont {B.}~\bibnamefont
  {Lorenz}}\ and\ \bibinfo {author} {\bibfnamefont {C.-W.}\ \bibnamefont
  {Chu}},\ }\bibinfo {title} {Superconducting ferromagnets: Ferromagnetic
  domains in the superconducting state},\
  \href{http://dx.doi.org/10.1038/nmat1423} {\bibfield  {journal} {\bibinfo
  {journal} {Nat. Mater.}\ }\textbf {\bibinfo {volume} {4}},\ \bibinfo {pages}
  {516} (\bibinfo {year} {2005})}\BibitemShut {NoStop}%
\bibitem [{not({\natexlab{a}})}]{note1}%
  \BibitemOpen
  \href@noop \ \bibinfo {note} {Note that the
  terminology ``ferromagnetic superconductor" is also occasionally employed in
  the literature related to the U-based germanides~\cite{Mineev}, which show
  {$T_\mathrm{m} > T_\mathrm{sc}$}.}\BibitemShut {Stop}%
\bibitem [{\citenamefont {Mineev}(2017)}]{Mineev}%
  \BibitemOpen
  \bibfield  {author} {\bibinfo {author} {\bibfnamefont {V.~P.}\ \bibnamefont
  {Mineev}},\ }\bibinfo {title} {Superconductivity in uranium ferromagnets},\
  \href{\doibase 10.3367/UFNe.2016.04.037771} {\bibfield  {journal} {\bibinfo
  {journal} {Physcis-Uspekhi}\ }\textbf {\bibinfo {volume} {60}},\ \bibinfo
  {pages} {121} (\bibinfo {year} {2017})}\BibitemShut {NoStop}%
\bibitem [{\citenamefont {Nachtrab}\ \emph {et~al.}(2006)\citenamefont
  {Nachtrab}, \citenamefont {Bernhard}, \citenamefont {Lin}, \citenamefont
  {Koelle},\ and\ \citenamefont {Kleiner}}]{Bernhard}%
  \BibitemOpen
  \bibfield  {author} {\bibinfo {author} {\bibfnamefont {T.}~\bibnamefont
  {Nachtrab}}, \bibinfo {author} {\bibfnamefont {C.}~\bibnamefont {Bernhard}},
  \bibinfo {author} {\bibfnamefont {C.}~\bibnamefont {Lin}}, \bibinfo {author}
  {\bibfnamefont {D.}~\bibnamefont {Koelle}}, \ and\ \bibinfo {author}
  {\bibfnamefont {R.}~\bibnamefont {Kleiner}},\ }\bibinfo {title} {The
  ruthenocuprates: natural superconductor-ferromagnet multilayers},\
  \href{\doibase 10.1016/j.crhy.2005.11.010} {\bibfield  {journal} {\bibinfo
  {journal} {Comptes Rendus Physique}\ }\textbf {\bibinfo {volume} {7}},\
  \bibinfo {pages} {68} (\bibinfo {year} {2006})}\BibitemShut {NoStop}%
\bibitem [{\citenamefont {Jiao}\ \emph {et~al.}(2017)\citenamefont {Jiao},
  \citenamefont {Tao}, \citenamefont {Ren}, \citenamefont {Liu},\ and\
  \citenamefont {Cao}}]{SVP.jiao}%
  \BibitemOpen
  \bibfield  {author} {\bibinfo {author} {\bibfnamefont {W.-H.}\ \bibnamefont
  {Jiao}}, \bibinfo {author} {\bibfnamefont {Q.}~\bibnamefont {Tao}}, \bibinfo
  {author} {\bibfnamefont {Z.}~\bibnamefont {Ren}}, \bibinfo {author}
  {\bibfnamefont {Y.}~\bibnamefont {Liu}}, \ and\ \bibinfo {author}
  {\bibfnamefont {G.-H.}\ \bibnamefont {Cao}},\ }\bibinfo {title} {Evidence of
  spontaneous vortex ground state in an iron-based ferromagnetic
  superconductor},\ \href{\doibase 10.1038/s41535-017-0057-0} {\bibfield
  {journal} {\bibinfo  {journal} {npj Quantum Materials}\ }\textbf {\bibinfo
  {volume} {2}},\ \bibinfo {pages} {50} (\bibinfo {year} {2017})}\BibitemShut
  {NoStop}%
\bibitem [{\citenamefont {Zapf}\ and\ \citenamefont {Dressel}(2017)}]{dressel}%
  \BibitemOpen
  \bibfield  {author} {\bibinfo {author} {\bibfnamefont {S.}~\bibnamefont
  {Zapf}}\ and\ \bibinfo {author} {\bibfnamefont {M.}~\bibnamefont {Dressel}},\
  }\bibinfo {title} {Europium-based iron pnictides: a unique laboratory for
  magnetism, superconductivity and structural effects},\
  \href{http://stacks.iop.org/0034-4885/80/i=1/a=016501} {\bibfield  {journal}
  {\bibinfo  {journal} {Rep. Prog. Phys.}\ }\textbf {\bibinfo {volume} {80}},\
  \bibinfo {pages} {016501} (\bibinfo {year} {2017})}\BibitemShut {NoStop}%
\bibitem [{\citenamefont {Ren}\ \emph {et~al.}(2008)\citenamefont {Ren},
  \citenamefont {Zhu}, \citenamefont {Jiang}, \citenamefont {Xu}, \citenamefont
  {Tao}, \citenamefont {Wang}, \citenamefont {Feng}, \citenamefont {Cao},\ and\
  \citenamefont {Xu}}]{ren2008}%
  \BibitemOpen
  \bibfield  {author} {\bibinfo {author} {\bibfnamefont {Z.}~\bibnamefont
  {Ren}}, \bibinfo {author} {\bibfnamefont {Z.}~\bibnamefont {Zhu}}, \bibinfo
  {author} {\bibfnamefont {S.}~\bibnamefont {Jiang}}, \bibinfo {author}
  {\bibfnamefont {X.}~\bibnamefont {Xu}}, \bibinfo {author} {\bibfnamefont
  {Q.}~\bibnamefont {Tao}}, \bibinfo {author} {\bibfnamefont {C.}~\bibnamefont
  {Wang}}, \bibinfo {author} {\bibfnamefont {C.}~\bibnamefont {Feng}}, \bibinfo
  {author} {\bibfnamefont {G.}~\bibnamefont {Cao}}, \ and\ \bibinfo {author}
  {\bibfnamefont {Z.}~\bibnamefont {Xu}},\ }\bibinfo {title} {Antiferromagnetic
  transition in ${\text{EuFe}}_{2}{\text{As}}_{2}$: A possible parent compound
  for superconductors},\ \href{\doibase 10.1103/PhysRevB.78.052501} {\bibfield
  {journal} {\bibinfo  {journal} {Phys. Rev. B}\ }\textbf {\bibinfo {volume}
  {78}},\ \bibinfo {pages} {052501} (\bibinfo {year} {2008})}\BibitemShut
  {NoStop}%
\bibitem [{\citenamefont {Jiang}\ \emph
  {et~al.}(2009{\natexlab{a}})\citenamefont {Jiang}, \citenamefont {Luo},
  \citenamefont {Ren}, \citenamefont {Zhu}, \citenamefont {Wang}, \citenamefont
  {Xu}, \citenamefont {Tao}, \citenamefont {Cao},\ and\ \citenamefont
  {Xu}}]{js-njp}%
  \BibitemOpen
  \bibfield  {author} {\bibinfo {author} {\bibfnamefont {S.}~\bibnamefont
  {Jiang}}, \bibinfo {author} {\bibfnamefont {Y.}~\bibnamefont {Luo}}, \bibinfo
  {author} {\bibfnamefont {Z.}~\bibnamefont {Ren}}, \bibinfo {author}
  {\bibfnamefont {Z.}~\bibnamefont {Zhu}}, \bibinfo {author} {\bibfnamefont
  {C.}~\bibnamefont {Wang}}, \bibinfo {author} {\bibfnamefont {X.}~\bibnamefont
  {Xu}}, \bibinfo {author} {\bibfnamefont {Q.}~\bibnamefont {Tao}}, \bibinfo
  {author} {\bibfnamefont {G.}~\bibnamefont {Cao}}, \ and\ \bibinfo {author}
  {\bibfnamefont {Z.}~\bibnamefont {Xu}},\ }\bibinfo {title} {Metamagnetic
  transition in ${\text{EuFe}}_{2}{\text{As}}_{2}$ single crystals},\
  \href{http://stacks.iop.org/1367-2630/11/i=2/a=025007} {\bibfield  {journal}
  {\bibinfo  {journal} {New J. Phys.}\ }\textbf {\bibinfo {volume} {11}},\
  \bibinfo {pages} {025007} (\bibinfo {year} {2009}{\natexlab{a}})}\BibitemShut
  {NoStop}%
\bibitem [{\citenamefont {Herrero-Mart\'{i}n}\ \emph
  {et~al.}(2009)\citenamefont {Herrero-Mart\'{i}n}, \citenamefont {Scagnoli},
  \citenamefont {Mazzoli}, \citenamefont {Su}, \citenamefont {Mittal},
  \citenamefont {Xiao}, \citenamefont {Brueckel}, \citenamefont {Kumar},
  \citenamefont {Dhar}, \citenamefont {Thamizhavel},\ and\ \citenamefont
  {Paolasini}}]{rxs2009}%
  \BibitemOpen
  \bibfield  {author} {\bibinfo {author} {\bibfnamefont {J.}~\bibnamefont
  {Herrero-Mart\'{i}n}}, \bibinfo {author} {\bibfnamefont {V.}~\bibnamefont
  {Scagnoli}}, \bibinfo {author} {\bibfnamefont {C.}~\bibnamefont {Mazzoli}},
  \bibinfo {author} {\bibfnamefont {Y.}~\bibnamefont {Su}}, \bibinfo {author}
  {\bibfnamefont {R.}~\bibnamefont {Mittal}}, \bibinfo {author} {\bibfnamefont
  {Y.}~\bibnamefont {Xiao}}, \bibinfo {author} {\bibfnamefont {T.}~\bibnamefont
  {Brueckel}}, \bibinfo {author} {\bibfnamefont {N.}~\bibnamefont {Kumar}},
  \bibinfo {author} {\bibfnamefont {S.~K.}\ \bibnamefont {Dhar}}, \bibinfo
  {author} {\bibfnamefont {A.}~\bibnamefont {Thamizhavel}}, \ and\ \bibinfo
  {author} {\bibfnamefont {L.}~\bibnamefont {Paolasini}},\ }\bibinfo {title}
  {Magnetic structure of ${\text{EuFe}}_{2}{\text{As}}_{2}$ as determined by
  resonant x-ray scattering},\ \href{\doibase 10.1103/PhysRevB.80.134411}
  {\bibfield  {journal} {\bibinfo  {journal} {Phys. Rev. B}\ }\textbf {\bibinfo
  {volume} {80}},\ \bibinfo {pages} {134411} (\bibinfo {year}
  {2009})}\BibitemShut {NoStop}%
\bibitem [{\citenamefont {Xiao}\ \emph {et~al.}(2009)\citenamefont {Xiao},
  \citenamefont {Su}, \citenamefont {Meven}, \citenamefont {Mittal},
  \citenamefont {Kumar}, \citenamefont {Chatterji}, \citenamefont {Price},
  \citenamefont {Persson}, \citenamefont {Kumar}, \citenamefont {Dhar},
  \citenamefont {Thamizhavel},\ and\ \citenamefont {Brueckel}}]{nd2009}%
  \BibitemOpen
  \bibfield  {author} {\bibinfo {author} {\bibfnamefont {Y.}~\bibnamefont
  {Xiao}}, \bibinfo {author} {\bibfnamefont {Y.}~\bibnamefont {Su}}, \bibinfo
  {author} {\bibfnamefont {M.}~\bibnamefont {Meven}}, \bibinfo {author}
  {\bibfnamefont {R.}~\bibnamefont {Mittal}}, \bibinfo {author} {\bibfnamefont
  {C.~M.~N.}\ \bibnamefont {Kumar}}, \bibinfo {author} {\bibfnamefont
  {T.}~\bibnamefont {Chatterji}}, \bibinfo {author} {\bibfnamefont
  {S.}~\bibnamefont {Price}}, \bibinfo {author} {\bibfnamefont
  {J.}~\bibnamefont {Persson}}, \bibinfo {author} {\bibfnamefont
  {N.}~\bibnamefont {Kumar}}, \bibinfo {author} {\bibfnamefont {S.~K.}\
  \bibnamefont {Dhar}}, \bibinfo {author} {\bibfnamefont {A.}~\bibnamefont
  {Thamizhavel}}, \ and\ \bibinfo {author} {\bibfnamefont {T.}~\bibnamefont
  {Brueckel}},\ }\bibinfo {title} {Magnetic structure of
  ${\text{EuFe}}_{2}{\text{As}}_{2}$ determined by single-crystal neutron
  diffraction},\ \href{\doibase 10.1103/PhysRevB.80.174424} {\bibfield
  {journal} {\bibinfo  {journal} {Phys. Rev. B}\ }\textbf {\bibinfo {volume}
  {80}},\ \bibinfo {pages} {174424} (\bibinfo {year} {2009})}\BibitemShut
  {NoStop}%
\bibitem [{\citenamefont {Ren}\ \emph {et~al.}(2009{\natexlab{a}})\citenamefont
  {Ren}, \citenamefont {Tao}, \citenamefont {Jiang}, \citenamefont {Feng},
  \citenamefont {Wang}, \citenamefont {Dai}, \citenamefont {Cao},\ and\
  \citenamefont {Xu}}]{ren2009}%
  \BibitemOpen
  \bibfield  {author} {\bibinfo {author} {\bibfnamefont {Z.}~\bibnamefont
  {Ren}}, \bibinfo {author} {\bibfnamefont {Q.}~\bibnamefont {Tao}}, \bibinfo
  {author} {\bibfnamefont {S.}~\bibnamefont {Jiang}}, \bibinfo {author}
  {\bibfnamefont {C.}~\bibnamefont {Feng}}, \bibinfo {author} {\bibfnamefont
  {C.}~\bibnamefont {Wang}}, \bibinfo {author} {\bibfnamefont {J.}~\bibnamefont
  {Dai}}, \bibinfo {author} {\bibfnamefont {G.}~\bibnamefont {Cao}}, \ and\
  \bibinfo {author} {\bibfnamefont {Z.}~\bibnamefont {Xu}},\ }\bibinfo {title}
  {Superconductivity induced by phosphorus doping and its coexistence with
  ferromagnetism in
  ${\mathrm{EuFe}}_{2}({\mathrm{As}}_{0.7}{\mathrm{P}}_{0.3}{)}_{2}$},\
  \href{\doibase 10.1103/PhysRevLett.102.137002} {\bibfield  {journal}
  {\bibinfo  {journal} {Phys. Rev. Lett.}\ }\textbf {\bibinfo {volume} {102}},\
  \bibinfo {pages} {137002} (\bibinfo {year} {2009}{\natexlab{a}})}\BibitemShut
  {NoStop}%
\bibitem [{\citenamefont {Jiang}\ \emph
  {et~al.}(2009{\natexlab{b}})\citenamefont {Jiang}, \citenamefont {Xing},
  \citenamefont {Xuan}, \citenamefont {Ren}, \citenamefont {Wang},
  \citenamefont {Xu},\ and\ \citenamefont {Cao}}]{jiang2009}%
  \BibitemOpen
  \bibfield  {author} {\bibinfo {author} {\bibfnamefont {S.}~\bibnamefont
  {Jiang}}, \bibinfo {author} {\bibfnamefont {H.}~\bibnamefont {Xing}},
  \bibinfo {author} {\bibfnamefont {G.}~\bibnamefont {Xuan}}, \bibinfo {author}
  {\bibfnamefont {Z.}~\bibnamefont {Ren}}, \bibinfo {author} {\bibfnamefont
  {C.}~\bibnamefont {Wang}}, \bibinfo {author} {\bibfnamefont {Z.}~\bibnamefont
  {Xu}}, \ and\ \bibinfo {author} {\bibfnamefont {G.}~\bibnamefont {Cao}},\
  }\bibinfo {title} {Superconductivity and local-moment magnetism in
  $\text{Eu}{({\text{Fe}}_{0.89}{\text{Co}}_{0.11})}_{2}{\text{As}}_{2}$},\
  \href{\doibase 10.1103/PhysRevB.80.184514} {\bibfield  {journal} {\bibinfo
  {journal} {Phys. Rev. B}\ }\textbf {\bibinfo {volume} {80}},\ \bibinfo
  {pages} {184514} (\bibinfo {year} {2009}{\natexlab{b}})}\BibitemShut
  {NoStop}%
\bibitem [{\citenamefont {Jiao}\ \emph {et~al.}(2011)\citenamefont {Jiao},
  \citenamefont {Tao}, \citenamefont {Bao}, \citenamefont {Sun}, \citenamefont
  {Feng}, \citenamefont {Xu}, \citenamefont {Nowik}, \citenamefont {Felner},\
  and\ \citenamefont {Cao}}]{jiao2011}%
  \BibitemOpen
  \bibfield  {author} {\bibinfo {author} {\bibfnamefont {W.-H.}\ \bibnamefont
  {Jiao}}, \bibinfo {author} {\bibfnamefont {Q.}~\bibnamefont {Tao}}, \bibinfo
  {author} {\bibfnamefont {J.-K.}\ \bibnamefont {Bao}}, \bibinfo {author}
  {\bibfnamefont {Y.-L.}\ \bibnamefont {Sun}}, \bibinfo {author} {\bibfnamefont
  {C.-M.}\ \bibnamefont {Feng}}, \bibinfo {author} {\bibfnamefont {Z.-A.}\
  \bibnamefont {Xu}}, \bibinfo {author} {\bibfnamefont {I.}~\bibnamefont
  {Nowik}}, \bibinfo {author} {\bibfnamefont {I.}~\bibnamefont {Felner}}, \
  and\ \bibinfo {author} {\bibfnamefont {G.-H.}\ \bibnamefont {Cao}},\
  }\bibinfo {title} {Anisotropic superconductivity in
  Eu(Fe$_{0.75}$Ru$_{0.25}$)$_2$As$_2$ ferromagnetic superconductor},\
  \href{http://stacks.iop.org/0295-5075/95/i=6/a=67007} {\bibfield  {journal}
  {\bibinfo  {journal} {EPL (Europhysics Letters)}\ }\textbf {\bibinfo {volume}
  {95}},\ \bibinfo {pages} {67007} (\bibinfo {year} {2011})}\BibitemShut
  {NoStop}%
\bibitem [{\citenamefont {Jiao}\ \emph {et~al.}(2013)\citenamefont {Jiao},
  \citenamefont {Zhai}, \citenamefont {Bao}, \citenamefont {Luo}, \citenamefont
  {Tao}, \citenamefont {Feng}, \citenamefont {Xu},\ and\ \citenamefont
  {Cao}}]{jiao2013}%
  \BibitemOpen
  \bibfield  {author} {\bibinfo {author} {\bibfnamefont {W.-H.}\ \bibnamefont
  {Jiao}}, \bibinfo {author} {\bibfnamefont {H.-F.}\ \bibnamefont {Zhai}},
  \bibinfo {author} {\bibfnamefont {J.-K.}\ \bibnamefont {Bao}}, \bibinfo
  {author} {\bibfnamefont {Y.-K.}\ \bibnamefont {Luo}}, \bibinfo {author}
  {\bibfnamefont {Q.}~\bibnamefont {Tao}}, \bibinfo {author} {\bibfnamefont
  {C.-M.}\ \bibnamefont {Feng}}, \bibinfo {author} {\bibfnamefont {Z.-A.}\
  \bibnamefont {Xu}}, \ and\ \bibinfo {author} {\bibfnamefont {G.-H.}\
  \bibnamefont {Cao}},\ }\bibinfo {title} {Anomalous critical fields and the
  absence of Meissner state in Eu(Fe$_{0.88}$Ir$_{0.12}$)$_2$As$_2$ crystals},\
  \href{http://stacks.iop.org/1367-2630/15/i=11/a=113002} {\bibfield  {journal}
  {\bibinfo  {journal} {New J. Phys.}\ }\textbf {\bibinfo {volume} {15}},\
  \bibinfo {pages} {113002} (\bibinfo {year} {2013})}\BibitemShut {NoStop}%
\bibitem [{\citenamefont {Paramanik}\ \emph {et~al.}(2013)\citenamefont
  {Paramanik}, \citenamefont {Das}, \citenamefont {Prasad},\ and\ \citenamefont
  {Hossain}}]{hossian2013}%
  \BibitemOpen
  \bibfield  {author} {\bibinfo {author} {\bibfnamefont {U.~B.}\ \bibnamefont
  {Paramanik}}, \bibinfo {author} {\bibfnamefont {D.}~\bibnamefont {Das}},
  \bibinfo {author} {\bibfnamefont {R.}~\bibnamefont {Prasad}}, \ and\ \bibinfo
  {author} {\bibfnamefont {Z.}~\bibnamefont {Hossain}},\ }\bibinfo {title}
  {Reentrant superconductivity in Eu(Fe$_{1-x}$Ir$_x$)$_2$As$_2$},\
  \href{http://stacks.iop.org/0953-8984/25/i=26/a=265701} {\bibfield  {journal}
  {\bibinfo  {journal} {J. Phys.: Condens. Matter}\ }\textbf {\bibinfo {volume}
  {25}},\ \bibinfo {pages} {265701} (\bibinfo {year} {2013})}\BibitemShut
  {NoStop}%
\bibitem [{\citenamefont {Nandi}\ \emph
  {et~al.}(2014{\natexlab{a}})\citenamefont {Nandi}, \citenamefont {Jin},
  \citenamefont {Xiao}, \citenamefont {Su}, \citenamefont {Price},
  \citenamefont {Shukla}, \citenamefont {Strempfer}, \citenamefont {Jeevan},
  \citenamefont {Gegenwart},\ and\ \citenamefont {Br\"uckel}}]{rxs2014}%
  \BibitemOpen
  \bibfield  {author} {\bibinfo {author} {\bibfnamefont {S.}~\bibnamefont
  {Nandi}}, \bibinfo {author} {\bibfnamefont {W.~T.}\ \bibnamefont {Jin}},
  \bibinfo {author} {\bibfnamefont {Y.}~\bibnamefont {Xiao}}, \bibinfo {author}
  {\bibfnamefont {Y.}~\bibnamefont {Su}}, \bibinfo {author} {\bibfnamefont
  {S.}~\bibnamefont {Price}}, \bibinfo {author} {\bibfnamefont {D.~K.}\
  \bibnamefont {Shukla}}, \bibinfo {author} {\bibfnamefont {J.}~\bibnamefont
  {Strempfer}}, \bibinfo {author} {\bibfnamefont {H.~S.}\ \bibnamefont
  {Jeevan}}, \bibinfo {author} {\bibfnamefont {P.}~\bibnamefont {Gegenwart}}, \
  and\ \bibinfo {author} {\bibfnamefont {T.}~\bibnamefont {Br\"uckel}},\
  }\bibinfo {title} {Coexistence of superconductivity and ferromagnetism in
  P-doped ${\text{EuFe}}_{2}{\mathrm{As}}_{2}$},\ \href{\doibase
  10.1103/PhysRevB.89.014512} {\bibfield  {journal} {\bibinfo  {journal} {Phys.
  Rev. B}\ }\textbf {\bibinfo {volume} {89}},\ \bibinfo {pages} {014512}
  (\bibinfo {year} {2014}{\natexlab{a}})}\BibitemShut {NoStop}%
\bibitem [{\citenamefont {Nandi}\ \emph
  {et~al.}(2014{\natexlab{b}})\citenamefont {Nandi}, \citenamefont {Jin},
  \citenamefont {Xiao}, \citenamefont {Su}, \citenamefont {Price},
  \citenamefont {Schmidt}, \citenamefont {Schmalzl}, \citenamefont {Chatterji},
  \citenamefont {Jeevan}, \citenamefont {Gegenwart},\ and\ \citenamefont
  {Br\"uckel}}]{nd2014}%
  \BibitemOpen
  \bibfield  {author} {\bibinfo {author} {\bibfnamefont {S.}~\bibnamefont
  {Nandi}}, \bibinfo {author} {\bibfnamefont {W.~T.}\ \bibnamefont {Jin}},
  \bibinfo {author} {\bibfnamefont {Y.}~\bibnamefont {Xiao}}, \bibinfo {author}
  {\bibfnamefont {Y.}~\bibnamefont {Su}}, \bibinfo {author} {\bibfnamefont
  {S.}~\bibnamefont {Price}}, \bibinfo {author} {\bibfnamefont
  {W.}~\bibnamefont {Schmidt}}, \bibinfo {author} {\bibfnamefont
  {K.}~\bibnamefont {Schmalzl}}, \bibinfo {author} {\bibfnamefont
  {T.}~\bibnamefont {Chatterji}}, \bibinfo {author} {\bibfnamefont {H.~S.}\
  \bibnamefont {Jeevan}}, \bibinfo {author} {\bibfnamefont {P.}~\bibnamefont
  {Gegenwart}}, \ and\ \bibinfo {author} {\bibfnamefont {T.}~\bibnamefont
  {Br\"uckel}},\ }\bibinfo {title} {Coexistence of ferromagnetism and
  superconductivity in iron based pnictides: a time resolved magnetooptical
  study},\ \href{\doibase 10.1103/PhysRevB.90.094407} {\bibfield  {journal}
  {\bibinfo  {journal} {Phys. Rev. B}\ }\textbf {\bibinfo {volume} {90}},\
  \bibinfo {pages} {094407} (\bibinfo {year} {2014}{\natexlab{b}})}\BibitemShut
  {NoStop}%
\bibitem [{\citenamefont {Jin}\ \emph {et~al.}(2013)\citenamefont {Jin},
  \citenamefont {Nandi}, \citenamefont {Xiao}, \citenamefont {Su},
  \citenamefont {Zaharko}, \citenamefont {Guguchia}, \citenamefont {Bukowski},
  \citenamefont {Price}, \citenamefont {Jiao}, \citenamefont {Cao},\ and\
  \citenamefont {Br\"uckel}}]{jin-Co}%
  \BibitemOpen
  \bibfield  {author} {\bibinfo {author} {\bibfnamefont {W.~T.}\ \bibnamefont
  {Jin}}, \bibinfo {author} {\bibfnamefont {S.}~\bibnamefont {Nandi}}, \bibinfo
  {author} {\bibfnamefont {Y.}~\bibnamefont {Xiao}}, \bibinfo {author}
  {\bibfnamefont {Y.}~\bibnamefont {Su}}, \bibinfo {author} {\bibfnamefont
  {O.}~\bibnamefont {Zaharko}}, \bibinfo {author} {\bibfnamefont
  {Z.}~\bibnamefont {Guguchia}}, \bibinfo {author} {\bibfnamefont
  {Z.}~\bibnamefont {Bukowski}}, \bibinfo {author} {\bibfnamefont
  {S.}~\bibnamefont {Price}}, \bibinfo {author} {\bibfnamefont {W.~H.}\
  \bibnamefont {Jiao}}, \bibinfo {author} {\bibfnamefont {G.~H.}\ \bibnamefont
  {Cao}}, \ and\ \bibinfo {author} {\bibfnamefont {T.}~\bibnamefont
  {Br\"uckel}},\ }\bibinfo {title} {Magnetic structure of superconducting
  Eu(Fe$_{0.82}$Co$_{0.18}$)$_{2}$As$_{2}$ as revealed by single-crystal
  neutron diffraction},\ \href{\doibase 10.1103/PhysRevB.88.214516} {\bibfield
  {journal} {\bibinfo  {journal} {Phys. Rev. B}\ }\textbf {\bibinfo {volume}
  {88}},\ \bibinfo {pages} {214516} (\bibinfo {year} {2013})}\BibitemShut
  {NoStop}%
\bibitem [{\citenamefont {Jin}\ \emph {et~al.}(2015)\citenamefont {Jin},
  \citenamefont {Li}, \citenamefont {Su}, \citenamefont {Nandi}, \citenamefont
  {Xiao}, \citenamefont {Jiao}, \citenamefont {Meven}, \citenamefont {Sazonov},
  \citenamefont {Feng}, \citenamefont {Chen}, \citenamefont {Ting},
  \citenamefont {Cao},\ and\ \citenamefont {Br\"uckel}}]{jin-Ir}%
  \BibitemOpen
  \bibfield  {author} {\bibinfo {author} {\bibfnamefont {W.~T.}\ \bibnamefont
  {Jin}}, \bibinfo {author} {\bibfnamefont {W.}~\bibnamefont {Li}}, \bibinfo
  {author} {\bibfnamefont {Y.}~\bibnamefont {Su}}, \bibinfo {author}
  {\bibfnamefont {S.}~\bibnamefont {Nandi}}, \bibinfo {author} {\bibfnamefont
  {Y.}~\bibnamefont {Xiao}}, \bibinfo {author} {\bibfnamefont {W.~H.}\
  \bibnamefont {Jiao}}, \bibinfo {author} {\bibfnamefont {M.}~\bibnamefont
  {Meven}}, \bibinfo {author} {\bibfnamefont {A.~P.}\ \bibnamefont {Sazonov}},
  \bibinfo {author} {\bibfnamefont {E.}~\bibnamefont {Feng}}, \bibinfo {author}
  {\bibfnamefont {Y.}~\bibnamefont {Chen}}, \bibinfo {author} {\bibfnamefont
  {C.~S.}\ \bibnamefont {Ting}}, \bibinfo {author} {\bibfnamefont {G.~H.}\
  \bibnamefont {Cao}}, \ and\ \bibinfo {author} {\bibfnamefont
  {T.}~\bibnamefont {Br\"uckel}},\ }\bibinfo {title} {Magnetic ground state of
  superconducting
  $\mathrm{Eu}(\mathrm{Fe}{}_{0.88}\mathrm{Ir}{}_{0.12}){}_{2}\mathrm{As}{}_{2}$:
  A combined neutron diffraction and first-principles calculation study},\
  \href{\doibase 10.1103/PhysRevB.91.064506} {\bibfield  {journal} {\bibinfo
  {journal} {Phys. Rev. B}\ }\textbf {\bibinfo {volume} {91}},\ \bibinfo
  {pages} {064506} (\bibinfo {year} {2015})}\BibitemShut {NoStop}%
\bibitem [{\citenamefont {Anand}\ \emph {et~al.}(2015)\citenamefont {Anand},
  \citenamefont {Adroja}, \citenamefont {Bhattacharyya}, \citenamefont
  {Paramanik}, \citenamefont {Manuel}, \citenamefont {Hillier}, \citenamefont
  {Khalyavin},\ and\ \citenamefont {Hossain}}]{adroja-Ir}%
  \BibitemOpen
  \bibfield  {author} {\bibinfo {author} {\bibfnamefont {V.~K.}\ \bibnamefont
  {Anand}}, \bibinfo {author} {\bibfnamefont {D.~T.}\ \bibnamefont {Adroja}},
  \bibinfo {author} {\bibfnamefont {A.}~\bibnamefont {Bhattacharyya}}, \bibinfo
  {author} {\bibfnamefont {U.~B.}\ \bibnamefont {Paramanik}}, \bibinfo {author}
  {\bibfnamefont {P.}~\bibnamefont {Manuel}}, \bibinfo {author} {\bibfnamefont
  {A.~D.}\ \bibnamefont {Hillier}}, \bibinfo {author} {\bibfnamefont
  {D.}~\bibnamefont {Khalyavin}}, \ and\ \bibinfo {author} {\bibfnamefont
  {Z.}~\bibnamefont {Hossain}},\ }\bibinfo {title}
  {$\ensuremath{\mu}\mathrm{SR}$ and neutron diffraction investigations on the
  reentrant ferromagnetic superconductor
  $\mathrm{Eu}({\mathrm{Fe}}_{0.86}{\mathrm{Ir}}_{0.14}){}_{2}{\mathrm{As}}_{2}$},\
  \href{\doibase 10.1103/PhysRevB.91.094427} {\bibfield  {journal} {\bibinfo
  {journal} {Phys. Rev. B}\ }\textbf {\bibinfo {volume} {91}},\ \bibinfo
  {pages} {094427} (\bibinfo {year} {2015})}\BibitemShut {NoStop}%
\bibitem [{\citenamefont {Cao}\ \emph {et~al.}(2011)\citenamefont {Cao},
  \citenamefont {Xu}, \citenamefont {Ren}, \citenamefont {Jiang}, \citenamefont
  {Feng},\ and\ \citenamefont {Xu}}]{cgh2011}%
  \BibitemOpen
  \bibfield  {author} {\bibinfo {author} {\bibfnamefont {G.}~\bibnamefont
  {Cao}}, \bibinfo {author} {\bibfnamefont {S.}~\bibnamefont {Xu}}, \bibinfo
  {author} {\bibfnamefont {Z.}~\bibnamefont {Ren}}, \bibinfo {author}
  {\bibfnamefont {S.}~\bibnamefont {Jiang}}, \bibinfo {author} {\bibfnamefont
  {C.}~\bibnamefont {Feng}}, \ and\ \bibinfo {author} {\bibfnamefont
  {Z.}~\bibnamefont {Xu}},\ }\bibinfo {title} {Superconductivity and
  ferromagnetism in EuFe$_2$(As$_{1-x}$P$_x$)$_2$},\
  \href{http://stacks.iop.org/0953-8984/23/i=46/a=464204} {\bibfield  {journal}
  {\bibinfo  {journal} {J. Phys.: Condens. Matter}\ }\textbf {\bibinfo {volume}
  {23}},\ \bibinfo {pages} {464204} (\bibinfo {year} {2011})}\BibitemShut
  {NoStop}%
\bibitem [{\citenamefont {Jeevan}\ \emph {et~al.}(2011)\citenamefont {Jeevan},
  \citenamefont {Kasinathan}, \citenamefont {Rosner},\ and\ \citenamefont
  {Gegenwart}}]{jeevan}%
  \BibitemOpen
  \bibfield  {author} {\bibinfo {author} {\bibfnamefont {H.~S.}\ \bibnamefont
  {Jeevan}}, \bibinfo {author} {\bibfnamefont {D.}~\bibnamefont {Kasinathan}},
  \bibinfo {author} {\bibfnamefont {H.}~\bibnamefont {Rosner}}, \ and\ \bibinfo
  {author} {\bibfnamefont {P.}~\bibnamefont {Gegenwart}},\ }\bibinfo {title}
  {Interplay of antiferromagnetism, ferromagnetism, and superconductivity in
  EuFe${}_{2}$(As${}_{1\ensuremath{-}x}$P${}_{x}$)${}_{2}$ single crystals},\
  \href{\doibase 10.1103/PhysRevB.83.054511} {\bibfield  {journal} {\bibinfo
  {journal} {Phys. Rev. B}\ }\textbf {\bibinfo {volume} {83}},\ \bibinfo
  {pages} {054511} (\bibinfo {year} {2011})}\BibitemShut {NoStop}%
\bibitem [{\citenamefont {Nicklas}\ \emph {et~al.}(2011)\citenamefont
  {Nicklas}, \citenamefont {Kumar}, \citenamefont {Lengyel}, \citenamefont
  {Schnelle},\ and\ \citenamefont {Leithe-Jasper}}]{nicklas}%
  \BibitemOpen
  \bibfield  {author} {\bibinfo {author} {\bibfnamefont {M.}~\bibnamefont
  {Nicklas}}, \bibinfo {author} {\bibfnamefont {M.}~\bibnamefont {Kumar}},
  \bibinfo {author} {\bibfnamefont {E.}~\bibnamefont {Lengyel}}, \bibinfo
  {author} {\bibfnamefont {W.}~\bibnamefont {Schnelle}}, \ and\ \bibinfo
  {author} {\bibfnamefont {A.}~\bibnamefont {Leithe-Jasper}},\ }\bibinfo
  {title} {Competition of local-moment ferromagnetism and superconductivity in
  Co-substituted {EuFe$_2$As$_2$}},\
  \href{http://stacks.iop.org/1742-6596/273/i=1/a=012101} {\bibfield  {journal}
  {\bibinfo  {journal} {J. Phys.: Conf. Ser.}\ }\textbf {\bibinfo {volume}
  {273}},\ \bibinfo {pages} {012101} (\bibinfo {year} {2011})}\BibitemShut
  {NoStop}%
\bibitem [{\citenamefont {Hu}\ \emph {et~al.}(2011)\citenamefont {Hu},
  \citenamefont {Bud'ko}, \citenamefont {Straszheim},\ and\ \citenamefont
  {Canfield}}]{SrEu122}%
  \BibitemOpen
  \bibfield  {author} {\bibinfo {author} {\bibfnamefont {R.}~\bibnamefont
  {Hu}}, \bibinfo {author} {\bibfnamefont {S.~L.}\ \bibnamefont {Bud'ko}},
  \bibinfo {author} {\bibfnamefont {W.~E.}\ \bibnamefont {Straszheim}}, \ and\
  \bibinfo {author} {\bibfnamefont {P.~C.}\ \bibnamefont {Canfield}},\
  }\bibinfo {title} {Phase diagram of superconductivity and antiferromagnetism
  in single crystals of
  Sr(Fe${}_{1\ensuremath{-}x}$Co${}_{x}$)${}_{2}$As${}_{2}$ and
  Sr${}_{1\ensuremath{-}y}$Eu${}_{y}$(Fe${}_{0.88}$Co${}_{0.12}$)${}_{2}$As${}_{2}$},\
  \href{\doibase 10.1103/PhysRevB.83.094520} {\bibfield  {journal} {\bibinfo
  {journal} {Phys. Rev. B}\ }\textbf {\bibinfo {volume} {83}},\ \bibinfo
  {pages} {094520} (\bibinfo {year} {2011})}\BibitemShut {NoStop}%
\bibitem [{\citenamefont {Ren}\ \emph {et~al.}(2009{\natexlab{b}})\citenamefont
  {Ren}, \citenamefont {Lin}, \citenamefont {Tao}, \citenamefont {Jiang},
  \citenamefont {Zhu}, \citenamefont {Wang}, \citenamefont {Cao},\ and\
  \citenamefont {Xu}}]{ren2009Ni}%
  \BibitemOpen
  \bibfield  {author} {\bibinfo {author} {\bibfnamefont {Z.}~\bibnamefont
  {Ren}}, \bibinfo {author} {\bibfnamefont {X.}~\bibnamefont {Lin}}, \bibinfo
  {author} {\bibfnamefont {Q.}~\bibnamefont {Tao}}, \bibinfo {author}
  {\bibfnamefont {S.}~\bibnamefont {Jiang}}, \bibinfo {author} {\bibfnamefont
  {Z.}~\bibnamefont {Zhu}}, \bibinfo {author} {\bibfnamefont {C.}~\bibnamefont
  {Wang}}, \bibinfo {author} {\bibfnamefont {G.}~\bibnamefont {Cao}}, \ and\
  \bibinfo {author} {\bibfnamefont {Z.}~\bibnamefont {Xu}},\ }\bibinfo {title}
  {Suppression of spin-density-wave transition and emergence of ferromagnetic
  ordering of ${\text{Eu}}^{2+}$ moments in
  ${\text{EuFe}}_{2\ensuremath{-}x}{\text{Ni}}_{x}{\text{As}}_{2}$},\
  \href{\doibase 10.1103/PhysRevB.79.094426} {\bibfield  {journal} {\bibinfo
  {journal} {Phys. Rev. B}\ }\textbf {\bibinfo {volume} {79}},\ \bibinfo
  {pages} {094426} (\bibinfo {year} {2009}{\natexlab{b}})}\BibitemShut
  {NoStop}%
\bibitem [{\citenamefont {Saha}\ \emph {et~al.}(2009)\citenamefont {Saha},
  \citenamefont {Butch}, \citenamefont {Kirshenbaum},\ and\ \citenamefont
  {Paglione}}]{Sr122Ni}%
  \BibitemOpen
  \bibfield  {author} {\bibinfo {author} {\bibfnamefont {S.~R.}\ \bibnamefont
  {Saha}}, \bibinfo {author} {\bibfnamefont {N.~P.}\ \bibnamefont {Butch}},
  \bibinfo {author} {\bibfnamefont {K.}~\bibnamefont {Kirshenbaum}}, \ and\
  \bibinfo {author} {\bibfnamefont {J.}~\bibnamefont {Paglione}},\ }\bibinfo
  {title} {Evolution of bulk superconductivity in
  ${\text{SrFe}}_{2}{\text{As}}_{2}$ with Ni substitution},\ \href{\doibase
  10.1103/PhysRevB.79.224519} {\bibfield  {journal} {\bibinfo  {journal} {Phys.
  Rev. B}\ }\textbf {\bibinfo {volume} {79}},\ \bibinfo {pages} {224519}
  (\bibinfo {year} {2009})}\BibitemShut {NoStop}%
\bibitem [{\citenamefont {Kawashima}\ \emph {et~al.}(2016)\citenamefont
  {Kawashima}, \citenamefont {Kinjo}, \citenamefont {Nishio}, \citenamefont
  {Ishida}, \citenamefont {Fujihisa}, \citenamefont {Gotoh}, \citenamefont
  {Kihou}, \citenamefont {Eisaki}, \citenamefont {Yoshida},\ and\ \citenamefont
  {Iyo}}]{Eu1144}%
  \BibitemOpen
  \bibfield  {author} {\bibinfo {author} {\bibfnamefont {K.}~\bibnamefont
  {Kawashima}}, \bibinfo {author} {\bibfnamefont {T.}~\bibnamefont {Kinjo}},
  \bibinfo {author} {\bibfnamefont {T.}~\bibnamefont {Nishio}}, \bibinfo
  {author} {\bibfnamefont {S.}~\bibnamefont {Ishida}}, \bibinfo {author}
  {\bibfnamefont {H.}~\bibnamefont {Fujihisa}}, \bibinfo {author}
  {\bibfnamefont {Y.}~\bibnamefont {Gotoh}}, \bibinfo {author} {\bibfnamefont
  {K.}~\bibnamefont {Kihou}}, \bibinfo {author} {\bibfnamefont
  {H.}~\bibnamefont {Eisaki}}, \bibinfo {author} {\bibfnamefont
  {Y.}~\bibnamefont {Yoshida}}, \ and\ \bibinfo {author} {\bibfnamefont
  {A.}~\bibnamefont {Iyo}},\ }\bibinfo {title} {Superconductivity in Fe-based
  compound Eu$A$Fe$_{4}{\mathrm{As}}_{4}$ ($A$ = Rb and Cs)},\ \href@noop {}
  {\bibfield  {journal} {\bibinfo  {journal} {J. Phys. Soc. Jpn.}\ }\textbf
  {\bibinfo {volume} {85}},\ \bibinfo {pages} {064710} (\bibinfo {year}
  {2016})}\BibitemShut {NoStop}%
\bibitem [{\citenamefont {Liu}\ \emph {et~al.}(2016{\natexlab{a}})\citenamefont
  {Liu}, \citenamefont {Liu}, \citenamefont {Tang}, \citenamefont {Jiang},
  \citenamefont {Wang}, \citenamefont {Ablimit}, \citenamefont {Jiao},
  \citenamefont {Tao}, \citenamefont {Feng}, \citenamefont {Xu},\ and\
  \citenamefont {Cao}}]{Rb1144.ly}%
  \BibitemOpen
  \bibfield  {author} {\bibinfo {author} {\bibfnamefont {Y.}~\bibnamefont
  {Liu}}, \bibinfo {author} {\bibfnamefont {Y.-B.}\ \bibnamefont {Liu}},
  \bibinfo {author} {\bibfnamefont {Z.-T.}\ \bibnamefont {Tang}}, \bibinfo
  {author} {\bibfnamefont {H.}~\bibnamefont {Jiang}}, \bibinfo {author}
  {\bibfnamefont {Z.-C.}\ \bibnamefont {Wang}}, \bibinfo {author}
  {\bibfnamefont {A.}~\bibnamefont {Ablimit}}, \bibinfo {author} {\bibfnamefont
  {W.-H.}\ \bibnamefont {Jiao}}, \bibinfo {author} {\bibfnamefont
  {Q.}~\bibnamefont {Tao}}, \bibinfo {author} {\bibfnamefont {C.-M.}\
  \bibnamefont {Feng}}, \bibinfo {author} {\bibfnamefont {Z.-A.}\ \bibnamefont
  {Xu}}, \ and\ \bibinfo {author} {\bibfnamefont {G.-H.}\ \bibnamefont {Cao}},\
  }\bibinfo {title} {Superconductivity and ferromagnetism in hole-doped
  ${\mathrm{RbEuFe}}_{4}{\mathrm{As}}_{4}$},\ \href{\doibase
  10.1103/PhysRevB.93.214503} {\bibfield  {journal} {\bibinfo  {journal} {Phys.
  Rev. B}\ }\textbf {\bibinfo {volume} {93}},\ \bibinfo {pages} {214503}
  (\bibinfo {year} {2016}{\natexlab{a}})}\BibitemShut {NoStop}%
\bibitem [{\citenamefont {Liu}\ \emph {et~al.}(2016{\natexlab{b}})\citenamefont
  {Liu}, \citenamefont {Liu}, \citenamefont {Chen}, \citenamefont {Tang},
  \citenamefont {Jiao}, \citenamefont {Tao}, \citenamefont {Xu},\ and\
  \citenamefont {Cao}}]{Cs1144.ly}%
  \BibitemOpen
  \bibfield  {author} {\bibinfo {author} {\bibfnamefont {Y.}~\bibnamefont
  {Liu}}, \bibinfo {author} {\bibfnamefont {Y.-B.}\ \bibnamefont {Liu}},
  \bibinfo {author} {\bibfnamefont {Q.}~\bibnamefont {Chen}}, \bibinfo {author}
  {\bibfnamefont {Z.-T.}\ \bibnamefont {Tang}}, \bibinfo {author}
  {\bibfnamefont {W.-H.}\ \bibnamefont {Jiao}}, \bibinfo {author}
  {\bibfnamefont {Q.}~\bibnamefont {Tao}}, \bibinfo {author} {\bibfnamefont
  {Z.-A.}\ \bibnamefont {Xu}}, \ and\ \bibinfo {author} {\bibfnamefont {G.-H.}\
  \bibnamefont {Cao}},\ }\bibinfo {title} {A new ferromagnetic superconductor:
  ${\mathrm{CsEuFe}}_{4}{\mathrm{As}}_{4}$},\ \href{\doibase
  10.1007/s11434-016-1139-2} {\bibfield  {journal} {\bibinfo  {journal} {Sci.
  Bull.}\ }\textbf {\bibinfo {volume} {61}},\ \bibinfo {pages} {1213} (\bibinfo
  {year} {2016}{\natexlab{b}})}\BibitemShut {NoStop}%
\bibitem [{\citenamefont {Jiang}\ \emph {et~al.}(2013)\citenamefont {Jiang},
  \citenamefont {Sun}, \citenamefont {Xu},\ and\ \citenamefont
  {Cao}}]{1144.jh}%
  \BibitemOpen
  \bibfield  {author} {\bibinfo {author} {\bibfnamefont {H.}~\bibnamefont
  {Jiang}}, \bibinfo {author} {\bibfnamefont {Y.-L.}\ \bibnamefont {Sun}},
  \bibinfo {author} {\bibfnamefont {Z.-A.}\ \bibnamefont {Xu}}, \ and\ \bibinfo
  {author} {\bibfnamefont {G.-H.}\ \bibnamefont {Cao}},\ }\bibinfo {title}
  {Crystal chemistry and structural design of iron-based superconductors},\
  \href@noop {} {\bibfield  {journal} {\bibinfo  {journal} {Chin. Phys. B}\
  }\textbf {\bibinfo {volume} {22}},\ \bibinfo {pages} {087410} (\bibinfo
  {year} {2013})}\BibitemShut {NoStop}%
\bibitem [{\citenamefont {Iyo}\ \emph {et~al.}(2016)\citenamefont {Iyo},
  \citenamefont {Kawashima}, \citenamefont {Kinjo}, \citenamefont {Nishio},
  \citenamefont {Ishida}, \citenamefont {Fujihisa}, \citenamefont {Gotoh},
  \citenamefont {Kihou}, \citenamefont {Eisaki},\ and\ \citenamefont
  {Yoshida}}]{1144}%
  \BibitemOpen
  \bibfield  {author} {\bibinfo {author} {\bibfnamefont {A.}~\bibnamefont
  {Iyo}}, \bibinfo {author} {\bibfnamefont {K.}~\bibnamefont {Kawashima}},
  \bibinfo {author} {\bibfnamefont {T.}~\bibnamefont {Kinjo}}, \bibinfo
  {author} {\bibfnamefont {T.}~\bibnamefont {Nishio}}, \bibinfo {author}
  {\bibfnamefont {S.}~\bibnamefont {Ishida}}, \bibinfo {author} {\bibfnamefont
  {H.}~\bibnamefont {Fujihisa}}, \bibinfo {author} {\bibfnamefont
  {Y.}~\bibnamefont {Gotoh}}, \bibinfo {author} {\bibfnamefont
  {K.}~\bibnamefont {Kihou}}, \bibinfo {author} {\bibfnamefont
  {H.}~\bibnamefont {Eisaki}}, \ and\ \bibinfo {author} {\bibfnamefont
  {Y.}~\bibnamefont {Yoshida}},\ }\bibinfo {title} {New-structure-type Fe-based
  superconductors: CaAFe$_4$As$_4$ (A = K, Rb, Cs) and SrAFe$_4$As$_4$ (A = Rb,
  Cs)},\ \href{\doibase 10.1021/jacs.5b12571} {\bibfield  {journal} {\bibinfo
  {journal} {J. Am. Chem. Soc.}\ }\textbf {\bibinfo {volume} {138}},\ \bibinfo
  {pages} {3410} (\bibinfo {year} {2016})}\BibitemShut {NoStop}%
\bibitem [{\citenamefont {Cao}\ \emph {et~al.}(2009)\citenamefont {Cao},
  \citenamefont {Jiang}, \citenamefont {Lin}, \citenamefont {Wang},
  \citenamefont {Li}, \citenamefont {Ren}, \citenamefont {Tao}, \citenamefont
  {Feng}, \citenamefont {Dai}, \citenamefont {Xu},\ and\ \citenamefont
  {Zhang}}]{cgh2009}%
  \BibitemOpen
  \bibfield  {author} {\bibinfo {author} {\bibfnamefont {G.}~\bibnamefont
  {Cao}}, \bibinfo {author} {\bibfnamefont {S.}~\bibnamefont {Jiang}}, \bibinfo
  {author} {\bibfnamefont {X.}~\bibnamefont {Lin}}, \bibinfo {author}
  {\bibfnamefont {C.}~\bibnamefont {Wang}}, \bibinfo {author} {\bibfnamefont
  {Y.}~\bibnamefont {Li}}, \bibinfo {author} {\bibfnamefont {Z.}~\bibnamefont
  {Ren}}, \bibinfo {author} {\bibfnamefont {Q.}~\bibnamefont {Tao}}, \bibinfo
  {author} {\bibfnamefont {C.}~\bibnamefont {Feng}}, \bibinfo {author}
  {\bibfnamefont {J.}~\bibnamefont {Dai}}, \bibinfo {author} {\bibfnamefont
  {Z.}~\bibnamefont {Xu}}, \ and\ \bibinfo {author} {\bibfnamefont {F.-C.}\
  \bibnamefont {Zhang}},\ }\bibinfo {title} {Narrow superconducting window in
  ${\text{LaFe}}_{1\ensuremath{-}x}{\text{Ni}}_{x}\text{AsO}$},\ \href{\doibase
  10.1103/PhysRevB.79.174505} {\bibfield  {journal} {\bibinfo  {journal} {Phys.
  Rev. B}\ }\textbf {\bibinfo {volume} {79}},\ \bibinfo {pages} {174505}
  (\bibinfo {year} {2009})}\BibitemShut {NoStop}%
\bibitem [{\citenamefont {Li}\ \emph {et~al.}(2009)\citenamefont {Li},
  \citenamefont {Luo}, \citenamefont {Wang}, \citenamefont {Chen},
  \citenamefont {Ren}, \citenamefont {Tao}, \citenamefont {Li}, \citenamefont
  {Lin}, \citenamefont {He}, \citenamefont {Zhu}, \citenamefont {Cao},\ and\
  \citenamefont {Xu}}]{llj}%
  \BibitemOpen
  \bibfield  {author} {\bibinfo {author} {\bibfnamefont {L.~J.}\ \bibnamefont
  {Li}}, \bibinfo {author} {\bibfnamefont {Y.~K.}\ \bibnamefont {Luo}},
  \bibinfo {author} {\bibfnamefont {Q.~B.}\ \bibnamefont {Wang}}, \bibinfo
  {author} {\bibfnamefont {H.}~\bibnamefont {Chen}}, \bibinfo {author}
  {\bibfnamefont {Z.}~\bibnamefont {Ren}}, \bibinfo {author} {\bibfnamefont
  {Q.}~\bibnamefont {Tao}}, \bibinfo {author} {\bibfnamefont {Y.~K.}\
  \bibnamefont {Li}}, \bibinfo {author} {\bibfnamefont {X.}~\bibnamefont
  {Lin}}, \bibinfo {author} {\bibfnamefont {M.}~\bibnamefont {He}}, \bibinfo
  {author} {\bibfnamefont {Z.~W.}\ \bibnamefont {Zhu}}, \bibinfo {author}
  {\bibfnamefont {G.~H.}\ \bibnamefont {Cao}}, \ and\ \bibinfo {author}
  {\bibfnamefont {Z.~A.}\ \bibnamefont {Xu}},\ }\bibinfo {title}
  {Superconductivity induced by Ni doping in {BaFe$_2$As$_2$} single
  crystals},\ \href{http://stacks.iop.org/1367-2630/11/i=2/a=025008} {\bibfield
   {journal} {\bibinfo  {journal} {New J. Phys.}\ }\textbf {\bibinfo {volume}
  {11}},\ \bibinfo {pages} {025008} (\bibinfo {year} {2009})}\BibitemShut
  {NoStop}%
\bibitem [{\citenamefont {Arrott}(1957)}]{Arrot}%
  \BibitemOpen
  \bibfield  {author} {\bibinfo {author} {\bibfnamefont {A.}~\bibnamefont
  {Arrott}},\ }\bibinfo {title} {Criterion for ferromagnetism from observations
  of magnetic isotherms},\ \href{\doibase 10.1103/PhysRev.108.1394} {\bibfield
  {journal} {\bibinfo  {journal} {Phys. Rev.}\ }\textbf {\bibinfo {volume}
  {108}},\ \bibinfo {pages} {1394} (\bibinfo {year} {1957})}\BibitemShut
  {NoStop}%
\bibitem [{\citenamefont {Zapf}\ \emph {et~al.}(2013)\citenamefont {Zapf},
  \citenamefont {Jeevan}, \citenamefont {Ivek}, \citenamefont {Pfister},
  \citenamefont {Klingert}, \citenamefont {Jiang}, \citenamefont {Wu},
  \citenamefont {Gegenwart}, \citenamefont {Kremer},\ and\ \citenamefont
  {Dressel}}]{zapf2013}%
  \BibitemOpen
  \bibfield  {author} {\bibinfo {author} {\bibfnamefont {S.}~\bibnamefont
  {Zapf}}, \bibinfo {author} {\bibfnamefont {H.~S.}\ \bibnamefont {Jeevan}},
  \bibinfo {author} {\bibfnamefont {T.}~\bibnamefont {Ivek}}, \bibinfo {author}
  {\bibfnamefont {F.}~\bibnamefont {Pfister}}, \bibinfo {author} {\bibfnamefont
  {F.}~\bibnamefont {Klingert}}, \bibinfo {author} {\bibfnamefont
  {S.}~\bibnamefont {Jiang}}, \bibinfo {author} {\bibfnamefont
  {D.}~\bibnamefont {Wu}}, \bibinfo {author} {\bibfnamefont {P.}~\bibnamefont
  {Gegenwart}}, \bibinfo {author} {\bibfnamefont {R.~K.}\ \bibnamefont
  {Kremer}}, \ and\ \bibinfo {author} {\bibfnamefont {M.}~\bibnamefont
  {Dressel}},\ }\bibinfo {title}
  {${\mathrm{EuFe}}_{2}({\mathrm{As}}_{1-x}{\mathrm{P}}_{x}{)}_{2}$: Reentrant
  spin glass and superconductivity},\ \href{\doibase
  10.1103/PhysRevLett.110.237002} {\bibfield  {journal} {\bibinfo  {journal}
  {Phys. Rev. Lett.}\ }\textbf {\bibinfo {volume} {110}},\ \bibinfo {pages}
  {237002} (\bibinfo {year} {2013})}\BibitemShut {NoStop}%
\bibitem [{\citenamefont {Anderson}\ and\ \citenamefont
  {Suhl}(1959)}]{Anderson}%
  \BibitemOpen
  \bibfield  {author} {\bibinfo {author} {\bibfnamefont {P.~W.}\ \bibnamefont
  {Anderson}}\ and\ \bibinfo {author} {\bibfnamefont {H.}~\bibnamefont
  {Suhl}},\ }\bibinfo {title} {Spin alignment in the superconducting state},\
  \href{\doibase 10.1103/PhysRev.116.898} {\bibfield  {journal} {\bibinfo
  {journal} {Phys. Rev.}\ }\textbf {\bibinfo {volume} {116}},\ \bibinfo {pages}
  {898} (\bibinfo {year} {1959})}\BibitemShut {NoStop}%
\bibitem [{\citenamefont {Faur\'e}\ and\ \citenamefont
  {Buzdin}(2005)}]{Buzdin}%
  \BibitemOpen
  \bibfield  {author} {\bibinfo {author} {\bibfnamefont {M.}~\bibnamefont
  {Faur\'e}}\ and\ \bibinfo {author} {\bibfnamefont {A.~I.}\ \bibnamefont
  {Buzdin}},\ }\bibinfo {title} {Domain structure in a superconducting
  ferromagnet},\ \href{\doibase 10.1103/PhysRevLett.94.187202} {\bibfield
  {journal} {\bibinfo  {journal} {Phys. Rev. Lett.}\ }\textbf {\bibinfo
  {volume} {94}},\ \bibinfo {pages} {187202} (\bibinfo {year}
  {2005})}\BibitemShut {NoStop}%
\bibitem [{\citenamefont {Veshchunov}\ \emph {et~al.}(2017)\citenamefont
  {Veshchunov}, \citenamefont {Vinnikov}, \citenamefont {Stolyarov},
  \citenamefont {Zhou}, \citenamefont {Shi}, \citenamefont {Xu}, \citenamefont
  {Grebenchuk}, \citenamefont {Baranov}, \citenamefont {Golovchanskiy},
  \citenamefont {Pyon}, \citenamefont {Sun}, \citenamefont {Jiao},
  \citenamefont {Cao}, \citenamefont {Tamegai},\ and\ \citenamefont
  {Golubov}}]{Eu122P.MFM}%
  \BibitemOpen
  \bibfield  {author} {\bibinfo {author} {\bibfnamefont {I.~S.}\ \bibnamefont
  {Veshchunov}}, \bibinfo {author} {\bibfnamefont {L.~Y.}\ \bibnamefont
  {Vinnikov}}, \bibinfo {author} {\bibfnamefont {V.~S.}\ \bibnamefont
  {Stolyarov}}, \bibinfo {author} {\bibfnamefont {N.}~\bibnamefont {Zhou}},
  \bibinfo {author} {\bibfnamefont {Z.~X.}\ \bibnamefont {Shi}}, \bibinfo
  {author} {\bibfnamefont {X.~F.}\ \bibnamefont {Xu}}, \bibinfo {author}
  {\bibfnamefont {S.~Y.}\ \bibnamefont {Grebenchuk}}, \bibinfo {author}
  {\bibfnamefont {D.~S.}\ \bibnamefont {Baranov}}, \bibinfo {author}
  {\bibfnamefont {I.~A.}\ \bibnamefont {Golovchanskiy}}, \bibinfo {author}
  {\bibfnamefont {S.}~\bibnamefont {Pyon}}, \bibinfo {author} {\bibfnamefont
  {Y.}~\bibnamefont {Sun}}, \bibinfo {author} {\bibfnamefont {W.}~\bibnamefont
  {Jiao}}, \bibinfo {author} {\bibfnamefont {G.}~\bibnamefont {Cao}}, \bibinfo
  {author} {\bibfnamefont {T.}~\bibnamefont {Tamegai}}, \ and\ \bibinfo
  {author} {\bibfnamefont {A.~A.}\ \bibnamefont {Golubov}},\ }\bibinfo {title}
  {Visualization of the magnetic flux structure in phosphorus-doped
  {EuFe$_2$As$_2$} single crystals},\ \href{\doibase 10.1134/s0021364017020151}
  {\bibfield  {journal} {\bibinfo  {journal} {JETP Letters}\ }\textbf {\bibinfo
  {volume} {105}},\ \bibinfo {pages} {98} (\bibinfo {year} {2017})}\BibitemShut
  {NoStop}%
\bibitem [{not({\natexlab{b}})}]{note2}%
  \BibitemOpen
  \href@noop \ \bibinfo{note}{Note that the
  experimental value of $M_{\mathrm{sat}}$ increases further at lower
  temperatures and at higher magnetic fields.}\BibitemShut {Stop}%
\bibitem [{not({\natexlab{c}})}]{note3}%
  \BibitemOpen
  \href@noop \ \bibinfo{note}{The internal magnetic
  field generated by Eu-spin ferromagnetism in RbEuFe$_4$As$_4$ is about a half
  of that ($\sim$9 kOe) for an Eu-based 122-type
  compound~\cite{jiao2013,SVP.jiao}, since the volume concentration of Eu
  moments of the former is only a half of that of the latter.}\BibitemShut
  {Stop}%
\bibitem [{\citenamefont {Tran}\ \emph {et~al.}(2012)\citenamefont {Tran},
  \citenamefont {Zaleski}, \citenamefont {Bukowski}, \citenamefont {Tran},\
  and\ \citenamefont {Zaleski}}]{Eu122Co.Tran}%
  \BibitemOpen
  \bibfield  {author} {\bibinfo {author} {\bibfnamefont {V.~H.}\ \bibnamefont
  {Tran}}, \bibinfo {author} {\bibfnamefont {T.~A.}\ \bibnamefont {Zaleski}},
  \bibinfo {author} {\bibfnamefont {Z.}~\bibnamefont {Bukowski}}, \bibinfo
  {author} {\bibfnamefont {L.~M.}\ \bibnamefont {Tran}}, \ and\ \bibinfo
  {author} {\bibfnamefont {A.~J.}\ \bibnamefont {Zaleski}},\ }\bibinfo {title}
  {Tuning superconductivity in {Eu(Fe$_{0.81}$Co$_{0.19}$)$_{2}$As$_{2}$} with
  magnetic fields},\ \href{\doibase 10.1103/PhysRevB.85.052502} {\bibfield
  {journal} {\bibinfo  {journal} {Phys. Rev. B}\ }\textbf {\bibinfo {volume}
  {85}},\ \bibinfo {pages} {052502} (\bibinfo {year} {2012})}\BibitemShut
  {NoStop}%
\bibitem [{\citenamefont {Chen}\ \emph {et~al.}(2009)\citenamefont {Chen},
  \citenamefont {Ren}, \citenamefont {Qiu}, \citenamefont {Bao}, \citenamefont
  {Liu}, \citenamefont {Wu}, \citenamefont {Wu}, \citenamefont {Xie},
  \citenamefont {Wang}, \citenamefont {Huang},\ and\ \citenamefont
  {Chen}}]{BaK122.cxh}%
  \BibitemOpen
  \bibfield  {author} {\bibinfo {author} {\bibfnamefont {H.}~\bibnamefont
  {Chen}}, \bibinfo {author} {\bibfnamefont {Y.}~\bibnamefont {Ren}}, \bibinfo
  {author} {\bibfnamefont {Y.}~\bibnamefont {Qiu}}, \bibinfo {author}
  {\bibfnamefont {W.}~\bibnamefont {Bao}}, \bibinfo {author} {\bibfnamefont
  {R.~H.}\ \bibnamefont {Liu}}, \bibinfo {author} {\bibfnamefont
  {G.}~\bibnamefont {Wu}}, \bibinfo {author} {\bibfnamefont {T.}~\bibnamefont
  {Wu}}, \bibinfo {author} {\bibfnamefont {Y.~L.}\ \bibnamefont {Xie}},
  \bibinfo {author} {\bibfnamefont {X.~F.}\ \bibnamefont {Wang}}, \bibinfo
  {author} {\bibfnamefont {Q.}~\bibnamefont {Huang}}, \ and\ \bibinfo {author}
  {\bibfnamefont {X.~H.}\ \bibnamefont {Chen}},\ }\bibinfo {title} {Coexistence
  of the spin-density wave and superconductivity in
  {Ba$_{1-x}$K$_x$Fe$_2$As$_2$}},\
  \href{http://stacks.iop.org/0295-5075/85/i=1/a=17006} {\bibfield  {journal}
  {\bibinfo  {journal} {EPL (Europhysics Letters)}\ }\textbf {\bibinfo {volume}
  {85}},\ \bibinfo {pages} {17006} (\bibinfo {year} {2009})}\BibitemShut
  {NoStop}%
\bibitem [{\citenamefont {Zinth}\ \emph {et~al.}(2011)\citenamefont {Zinth},
  \citenamefont {Dellmann}, \citenamefont {Klauss},\ and\ \citenamefont
  {Johrendt}}]{BaK122Co}%
  \BibitemOpen
  \bibfield  {author} {\bibinfo {author} {\bibfnamefont {V.}~\bibnamefont
  {Zinth}}, \bibinfo {author} {\bibfnamefont {T.}~\bibnamefont {Dellmann}},
  \bibinfo {author} {\bibfnamefont {H.-H.}\ \bibnamefont {Klauss}}, \ and\
  \bibinfo {author} {\bibfnamefont {D.}~\bibnamefont {Johrendt}},\ }\bibinfo
  {title} {Recovery of a parentlike state in
  {Ba$_{1-x}$K$_x$Fe$_{1.86}$Co$_{0.14}$As$_2$}},\ \href{\doibase
  10.1002/anie.201102866} {\bibfield  {journal} {\bibinfo  {journal} {Angew.
  Chem. Inter. Ed.}\ }\textbf {\bibinfo {volume} {50}},\ \bibinfo {pages}
  {7919} (\bibinfo {year} {2011})}\BibitemShut {NoStop}%
\bibitem [{\citenamefont {Anupam}\ \emph {et~al.}(2012)\citenamefont {Anupam},
  \citenamefont {Anand}, \citenamefont {Paulose}, \citenamefont {Ramakrishnan},
  \citenamefont {Geibel},\ and\ \citenamefont {Hossain}}]{EuK122Ni}%
  \BibitemOpen
  \bibfield  {author} {\bibinfo {author} {\bibnamefont {Anupam}}, \bibinfo
  {author} {\bibfnamefont {V.~K.}\ \bibnamefont {Anand}}, \bibinfo {author}
  {\bibfnamefont {P.~L.}\ \bibnamefont {Paulose}}, \bibinfo {author}
  {\bibfnamefont {S.}~\bibnamefont {Ramakrishnan}}, \bibinfo {author}
  {\bibfnamefont {C.}~\bibnamefont {Geibel}}, \ and\ \bibinfo {author}
  {\bibfnamefont {Z.}~\bibnamefont {Hossain}},\ }\bibinfo {title} {Effect of Ni
  doping on magnetism and superconductivity in
  {Eu$_{0.5}$K$_{0.5}$Fe$_2$As$_2$}},\ \href{<Go to ISI>://WOS:000302611400005}
  {\bibfield  {journal} {\bibinfo  {journal} {Phys. Rev. B}\ }\textbf {\bibinfo
  {volume} {85}} (\bibinfo {year} {2012})}\BibitemShut {NoStop}%
\bibitem [{\citenamefont {Meier}\ \emph {et~al.}(2017)\citenamefont {Meier},
  \citenamefont {Ding}, \citenamefont {Kreyssig}, \citenamefont {Bud'ko},
  \citenamefont {Sapkota}, \citenamefont {Kothapalli}, \citenamefont {Borisov},
  \citenamefont {Valenti}, \citenamefont {Batista}, \citenamefont {Orth},
  \citenamefont {Fernandes}, \citenamefont {Goldman}, \citenamefont {Furukawa},
  \citenamefont {B\"{o}hmer},\ and\ \citenamefont
  {Canfield}}]{1144Ni.canfield}%
  \BibitemOpen
  \bibfield  {author} {\bibinfo {author} {\bibfnamefont {W.~R.}\ \bibnamefont
  {Meier}}, \bibinfo {author} {\bibfnamefont {Q.-P.}\ \bibnamefont {Ding}},
  \bibinfo {author} {\bibfnamefont {A.}~\bibnamefont {Kreyssig}}, \bibinfo
  {author} {\bibfnamefont {S.~L.}\ \bibnamefont {Bud'ko}}, \bibinfo {author}
  {\bibfnamefont {A.}~\bibnamefont {Sapkota}}, \bibinfo {author} {\bibfnamefont
  {K.}~\bibnamefont {Kothapalli}}, \bibinfo {author} {\bibfnamefont
  {V.}~\bibnamefont {Borisov}}, \bibinfo {author} {\bibfnamefont
  {R.}~\bibnamefont {Valenti}}, \bibinfo {author} {\bibfnamefont {C.~D.}\
  \bibnamefont {Batista}}, \bibinfo {author} {\bibfnamefont {P.~P.}\
  \bibnamefont {Orth}}, \bibinfo {author} {\bibfnamefont {R.~M.}\ \bibnamefont
  {Fernandes}}, \bibinfo {author} {\bibfnamefont {A.~I.}\ \bibnamefont
  {Goldman}}, \bibinfo {author} {\bibfnamefont {Y.}~\bibnamefont {Furukawa}},
  \bibinfo {author} {\bibfnamefont {A.~E.}\ \bibnamefont {B\"{o}hmer}}, \ and\
  \bibinfo {author} {\bibfnamefont {P.~C.}\ \bibnamefont {Canfield}},\
  }\bibinfo {title} {Hedgehog spin vortex crystal in a hole-doped iron based
  superconductor},\ \href@noop {} {\bibfield  {journal} {\bibinfo  {journal}
  {arXiv}\ }\textbf {\bibinfo {volume} {1706}},\ \bibinfo {pages} {01067}
  (\bibinfo {year} {2017})}\BibitemShut {NoStop}%
\bibitem [{\citenamefont {Hardy}\ \emph {et~al.}(2016)\citenamefont {Hardy},
  \citenamefont {B\"ohmer}, \citenamefont {de' Medici}, \citenamefont {Capone},
  \citenamefont {Giovannetti}, \citenamefont {Eder}, \citenamefont {Wang},
  \citenamefont {He}, \citenamefont {Wolf}, \citenamefont {Schweiss},
  \citenamefont {Heid}, \citenamefont {Herbig}, \citenamefont {Adelmann},
  \citenamefont {Fisher},\ and\ \citenamefont {Meingast}}]{BaK122.HC}%
  \BibitemOpen
  \bibfield  {author} {\bibinfo {author} {\bibfnamefont {F.}~\bibnamefont
  {Hardy}}, \bibinfo {author} {\bibfnamefont {A.~E.}\ \bibnamefont {B\"ohmer}},
  \bibinfo {author} {\bibfnamefont {L.}~\bibnamefont {de' Medici}}, \bibinfo
  {author} {\bibfnamefont {M.}~\bibnamefont {Capone}}, \bibinfo {author}
  {\bibfnamefont {G.}~\bibnamefont {Giovannetti}}, \bibinfo {author}
  {\bibfnamefont {R.}~\bibnamefont {Eder}}, \bibinfo {author} {\bibfnamefont
  {L.}~\bibnamefont {Wang}}, \bibinfo {author} {\bibfnamefont {M.}~\bibnamefont
  {He}}, \bibinfo {author} {\bibfnamefont {T.}~\bibnamefont {Wolf}}, \bibinfo
  {author} {\bibfnamefont {P.}~\bibnamefont {Schweiss}}, \bibinfo {author}
  {\bibfnamefont {R.}~\bibnamefont {Heid}}, \bibinfo {author} {\bibfnamefont
  {A.}~\bibnamefont {Herbig}}, \bibinfo {author} {\bibfnamefont
  {P.}~\bibnamefont {Adelmann}}, \bibinfo {author} {\bibfnamefont {R.~A.}\
  \bibnamefont {Fisher}}, \ and\ \bibinfo {author} {\bibfnamefont
  {C.}~\bibnamefont {Meingast}},\ }\bibinfo {title} {Strong correlations,
  strong coupling, and $s$-wave superconductivity in hole-doped
  ${\mathrm{BaFe}}_{2}{\mathrm{As}}_{2}$ single crystals},\ \href{\doibase
  10.1103/PhysRevB.94.205113} {\bibfield  {journal} {\bibinfo  {journal} {Phys.
  Rev. B}\ }\textbf {\bibinfo {volume} {94}},\ \bibinfo {pages} {205113}
  (\bibinfo {year} {2016})}\BibitemShut {NoStop}%
\bibitem [{\citenamefont {Holmes}\ and\ \citenamefont
  {Schieber}(1966)}]{Eu3O4}%
  \BibitemOpen
  \bibfield  {author} {\bibinfo {author} {\bibfnamefont {L.}~\bibnamefont
  {Holmes}}\ and\ \bibinfo {author} {\bibfnamefont {M.}~\bibnamefont
  {Schieber}},\ }\bibinfo {title} {Magnetic Ordering in Eu$_3$O$_4$ and
  EuGd$_2$O$_4$},\ \href{\doibase 10.1063/1.1708542} {\bibfield  {journal}
  {\bibinfo  {journal} {Journal of Applied Physics}\ }\textbf {\bibinfo
  {volume} {37}},\ \bibinfo {pages} {968} (\bibinfo {year} {1966})}\BibitemShut
  {NoStop}%
\bibitem [{\citenamefont {Avci}\ \emph {et~al.}(2012)\citenamefont {Avci},
  \citenamefont {Chmaissem}, \citenamefont {Chung}, \citenamefont {Rosenkranz},
  \citenamefont {Goremychkin}, \citenamefont {Castellan}, \citenamefont
  {Todorov}, \citenamefont {Schlueter}, \citenamefont {Claus}, \citenamefont
  {Daoud-Aladine}, \citenamefont {Khalyavin}, \citenamefont {Kanatzidis},\ and\
  \citenamefont {Osborn}}]{BaK122.Kanatzidis}%
  \BibitemOpen
  \bibfield  {author} {\bibinfo {author} {\bibfnamefont {S.}~\bibnamefont
  {Avci}}, \bibinfo {author} {\bibfnamefont {O.}~\bibnamefont {Chmaissem}},
  \bibinfo {author} {\bibfnamefont {D.~Y.}\ \bibnamefont {Chung}}, \bibinfo
  {author} {\bibfnamefont {S.}~\bibnamefont {Rosenkranz}}, \bibinfo {author}
  {\bibfnamefont {E.~A.}\ \bibnamefont {Goremychkin}}, \bibinfo {author}
  {\bibfnamefont {J.~P.}\ \bibnamefont {Castellan}}, \bibinfo {author}
  {\bibfnamefont {I.~S.}\ \bibnamefont {Todorov}}, \bibinfo {author}
  {\bibfnamefont {J.~A.}\ \bibnamefont {Schlueter}}, \bibinfo {author}
  {\bibfnamefont {H.}~\bibnamefont {Claus}}, \bibinfo {author} {\bibfnamefont
  {A.}~\bibnamefont {Daoud-Aladine}}, \bibinfo {author} {\bibfnamefont {D.~D.}\
  \bibnamefont {Khalyavin}}, \bibinfo {author} {\bibfnamefont {M.~G.}\
  \bibnamefont {Kanatzidis}}, \ and\ \bibinfo {author} {\bibfnamefont
  {R.}~\bibnamefont {Osborn}},\ }\bibinfo {title} {Phase diagram of
  Ba${}_{1\ensuremath{-}x}$K${}_{x}$Fe${}_{2}$As${}_{2}$},\ \href{\doibase
  10.1103/PhysRevB.85.184507} {\bibfield  {journal} {\bibinfo  {journal} {Phys.
  Rev. B}\ }\textbf {\bibinfo {volume} {85}},\ \bibinfo {pages} {184507}
  (\bibinfo {year} {2012})}\BibitemShut {NoStop}%
\bibitem [{\citenamefont {Jin}\ \emph {et~al.}(2016)\citenamefont {Jin},
  \citenamefont {Xiao}, \citenamefont {Bukowski}, \citenamefont {Su},
  \citenamefont {Nandi}, \citenamefont {Sazonov}, \citenamefont {Meven},
  \citenamefont {Zaharko}, \citenamefont {Demirdis}, \citenamefont {Nemkovski},
  \citenamefont {Schmalzl}, \citenamefont {Tran}, \citenamefont {Guguchia},
  \citenamefont {Feng}, \citenamefont {Fu},\ and\ \citenamefont
  {Br\"uckel}}]{Eu122Co.jwt}%
  \BibitemOpen
  \bibfield  {author} {\bibinfo {author} {\bibfnamefont {W.~T.}\ \bibnamefont
  {Jin}}, \bibinfo {author} {\bibfnamefont {Y.}~\bibnamefont {Xiao}}, \bibinfo
  {author} {\bibfnamefont {Z.}~\bibnamefont {Bukowski}}, \bibinfo {author}
  {\bibfnamefont {Y.}~\bibnamefont {Su}}, \bibinfo {author} {\bibfnamefont
  {S.}~\bibnamefont {Nandi}}, \bibinfo {author} {\bibfnamefont {A.~P.}\
  \bibnamefont {Sazonov}}, \bibinfo {author} {\bibfnamefont {M.}~\bibnamefont
  {Meven}}, \bibinfo {author} {\bibfnamefont {O.}~\bibnamefont {Zaharko}},
  \bibinfo {author} {\bibfnamefont {S.}~\bibnamefont {Demirdis}}, \bibinfo
  {author} {\bibfnamefont {K.}~\bibnamefont {Nemkovski}}, \bibinfo {author}
  {\bibfnamefont {K.}~\bibnamefont {Schmalzl}}, \bibinfo {author}
  {\bibfnamefont {L.~M.}\ \bibnamefont {Tran}}, \bibinfo {author}
  {\bibfnamefont {Z.}~\bibnamefont {Guguchia}}, \bibinfo {author}
  {\bibfnamefont {E.}~\bibnamefont {Feng}}, \bibinfo {author} {\bibfnamefont
  {Z.}~\bibnamefont {Fu}}, \ and\ \bibinfo {author} {\bibfnamefont
  {T.}~\bibnamefont {Br\"uckel}},\ }\bibinfo {title} {Phase diagram of Eu
  magnetic ordering in Sn-flux-grown {Eu(Fe$_{1-x}$Co$_{x}$)$_{2}$As$_{2}$}
  single crystals},\ \href{\doibase 10.1103/PhysRevB.94.184513} {\bibfield
  {journal} {\bibinfo  {journal} {Phys. Rev. B}\ }\textbf {\bibinfo {volume}
  {94}},\ \bibinfo {pages} {184513} (\bibinfo {year} {2016})}\BibitemShut
  {NoStop}%
\bibitem [{\citenamefont {Su}\ \emph {et~al.}(2009)\citenamefont {Su},
  \citenamefont {Link}, \citenamefont {Schneidewind}, \citenamefont {Wolf},
  \citenamefont {Adelmann}, \citenamefont {Xiao}, \citenamefont {Meven},
  \citenamefont {Mittal}, \citenamefont {Rotter}, \citenamefont {Johrendt},
  \citenamefont {Brueckel},\ and\ \citenamefont {Loewenhaupt}}]{Sn-flux}%
  \BibitemOpen
  \bibfield  {author} {\bibinfo {author} {\bibfnamefont {Y.}~\bibnamefont
  {Su}}, \bibinfo {author} {\bibfnamefont {P.}~\bibnamefont {Link}}, \bibinfo
  {author} {\bibfnamefont {A.}~\bibnamefont {Schneidewind}}, \bibinfo {author}
  {\bibfnamefont {T.}~\bibnamefont {Wolf}}, \bibinfo {author} {\bibfnamefont
  {P.}~\bibnamefont {Adelmann}}, \bibinfo {author} {\bibfnamefont
  {Y.}~\bibnamefont {Xiao}}, \bibinfo {author} {\bibfnamefont {M.}~\bibnamefont
  {Meven}}, \bibinfo {author} {\bibfnamefont {R.}~\bibnamefont {Mittal}},
  \bibinfo {author} {\bibfnamefont {M.}~\bibnamefont {Rotter}}, \bibinfo
  {author} {\bibfnamefont {D.}~\bibnamefont {Johrendt}}, \bibinfo {author}
  {\bibfnamefont {T.}~\bibnamefont {Brueckel}}, \ and\ \bibinfo {author}
  {\bibfnamefont {M.}~\bibnamefont {Loewenhaupt}},\ }\bibinfo {title}
  {Antiferromagnetic ordering and structural phase transition in
  ${\text{Ba}}_{2}{\text{Fe}}_{2}{\text{As}}_{2}$ with Sn incorporated from the
  growth flux},\ \href{\doibase 10.1103/PhysRevB.79.064504} {\bibfield
  {journal} {\bibinfo  {journal} {Phys. Rev. B}\ }\textbf {\bibinfo {volume}
  {79}},\ \bibinfo {pages} {064504} (\bibinfo {year} {2009})}\BibitemShut
  {NoStop}%
\bibitem [{\citenamefont {Passell}\ \emph {et~al.}(1976)\citenamefont
  {Passell}, \citenamefont {Dietrich},\ and\ \citenamefont
  {Als-Nielsen}}]{EuCh.neutron}%
  \BibitemOpen
  \bibfield  {author} {\bibinfo {author} {\bibfnamefont {L.}~\bibnamefont
  {Passell}}, \bibinfo {author} {\bibfnamefont {O.~W.}\ \bibnamefont
  {Dietrich}}, \ and\ \bibinfo {author} {\bibfnamefont {J.}~\bibnamefont
  {Als-Nielsen}},\ }\bibinfo {title} {Neutron scattering from the Heisenberg
  ferromagnets EuO and EuS. I. The exchange interactions},\
  \href{https://link.aps.org/doi/10.1103/PhysRevB.14.4897} {\bibfield
  {journal} {\bibinfo  {journal} {Phys. Rev. B}\ }\textbf {\bibinfo {volume}
  {14}},\ \bibinfo {pages} {4897} (\bibinfo {year} {1976})}\BibitemShut
  {NoStop}%
\bibitem [{\citenamefont {T.}(1970)}]{Kasuya}%
  \BibitemOpen
  \bibfield  {author} {\bibinfo {author} {\bibfnamefont {K.}~\bibnamefont
  {T.}},\ }\bibinfo {title} {Exchange mechanisms in Europium chalcogenides},\
  \href@noop {} {\bibfield  {journal} {\bibinfo  {journal} {IBM J. Res. Dev.}\
  }\textbf {\bibinfo {volume} {14}},\ \bibinfo {pages} {214} (\bibinfo {year}
  {1970})}\BibitemShut {NoStop}%
\end{thebibliography}
%

\end{document}